# Strain-Driven Terahertz Phase Modulation in Piezoelectric and Nonlinear Optical Crystals: Mechanisms, Material Developments, and Modulation Strategies

Aarifa Muhammed Ashraf [a], and Rajaboopathi Mani[a*]

[a]Key Laboratory of Emergent Materials, Department of Physics and Nanotechnology, Faculty of Engineering and Technology, SRM Institute of Science and Technology, Kattankulathur-603203, Chengalpattu Dt, Tamil Nadu, India.

[*]Corresponding author: rajaboom@srmist.edu.in

## Abstract

Terahertz (THz) technology has witnessed notable developments in sensing, spectroscopy, imaging, and wireless communication which have increased the demand for high-performance THz modulators. Among the various approaches for controlling the THz waves, external stimuli such as optical pumping, thermal and electric fields on materials are widely employed as promising strategy for modulation. This review focuses on recent advances in THz modulation based on different materials including 2D semiconductors and metamaterials with attention to their response under external stimuli. In addition, strain-driven THz modulation in non-centrosymmetric piezoelectric materials including organic and semi-organic crystals is examined with emphasis on their ability to establish strong coupling between propagating THz waves and lattice vibrations under piezoelectric resonance conditions. We further highlight earlier reports on strain-driven THz modulation in lithium niobate ($LiNbO_3$) and critically assessed the potential of several other nonlinear optical crystals exhibiting piezoelectric resonance peaks suitable for strain-driven THz modulation. An overview of THz modulators integrated with tunable materials is presented highlighting their operating principles and key performance metrics comprising modulation speed, bandwidth, modulation depth and insertion loss.

## 1. Introduction

The nonlinear optical (NLO) materials have gained immense focus in various fields due to the observation of second harmonic generation (SHG) and other nonlinear phenomena such as optical rectification, Kerr effect and two photon absorption. [1,2] A wide range of inorganic crystals like ZnTe, GaP and $LiNbO_3$ are in use, but organic crystals have resulted in non-linear optical phenomena with high SHG, broadband terahertz generation, multi-phonon imaging and quantum optics.[3] Recently, NLO frequency conversion of THz waves has become a significant approach for both generating and modulating the THz signals. Enhancing the conversion efficiency in THz regime holds a potential importance in the design and optimization of THz modulation for next generation THz communications.[4]

THz region in electromagnetic spectrum spanning frequencies from 0.1 to 10 THz, exhibiting a promising application in the field of imaging, radar detection, sensing, security checking, wireless communications, and spectroscopy.[5–7] THz signal can be generated mainly by laser using different methods such as optical rectification in NLO crystals, photoconductive antennas, plasma sources, and laser-induced acoustic vibrations. In optical rectification, femtosecond pulsed laser will be exposed to non-centrosymmetric crystals like ZnTe, $LiNbO_3$, and organic DAST and its derivatives to produce broadband THz output. The conversion efficiency achieved through this method is often low around $10^{-4}$ to $10^{-6}$ and phase-matching is a persistent challenge.[8] Photoconductive antennas (PCAs) generate THz pulses by accelerating photoexcited carriers under bias and they are widely used due to simplified semiconductor material design.[9] However, scaling to higher power and maintaining broadband (< 5 THz) response is based on power of input laser. Recently, THz generation has been achieved using acoustic methods where high-frequency mechanical vibrations in mesocrystals generate THz.[8,10] The most effective way to create high-intensity THz fields with incredibly wide bandwidths among the other THz wave generation techniques is optical rectification of near-infrared (IR) rays using organic NLO crystals. In communication, THz radiation has achieved 10 Gbps of transmission speed opted to use for 6G network which is 100 to 1000 times faster than the currently used ultra-wideband (UWB) technologies. THz communication exhibits the combined advantages over microwave and infrared systems, showing ~1-4 orders larger communication capacity than the microwave links. [11] Technology such as modulation and demodulation, THz detection, and THz wave generation are crucial for THz communication.[12] In order to implement the THz waves into practical applications, it is necessary to develop effective THz modulators and switches that enable complete control and manipulation of THz waves. Modulators control key properties of electromagnetic waves such as amplitude, phase, temporal and spatial characteristics which are classified based on the controlled parameter or the modulation technique.[13] Using techniques like phase-shift keying (PSK) or quadrature amplitude modulation (QAM), information can be encoded onto THz carrier waves. Beam steering is the controlled redirection of EM signal in modulation. Fast and low-loss modulators are essential for achieving the data rates needed beyond 5G and 6G networks. By introducing spatial phase gradients across a metasurface or phase array, THz beams can be directed without mechanical movement.

Recent studies have reported various THz modulators[14,15] employing various materials and structural design to achieve the crucial parameters of modulation including modulation depth, modulation speed, bandwidth, and insertion losses. Several studies have reported high modulation performance, demonstrating phase modulation ranging from 36° to 360° with device loss below 3dB. Modulation techniques based on free-space terahertz phase modulator (FSPM) and guided wave terahertz phase modulator (GWPM) with the integration of varying material design like graphene, liquid crystals, semiconductors, phase change materials, meta-chip phase modulators are well explored. In FSPM, the multistage phase modulating layers are used to achieve large phase shift. However, the increased number of layers causes high insertion loss. Whereas GWPM confines the THz wave within a waveguide, enabling efficient phase modulation with comparatively low insertion loss.[16] Introducing strain into the materials operated at the piezoelectric resonance and thus change in refractive index is a new approach for THz modulation which enables strong coupling between lattice phonons and polarization gradients. It controls the incident THz waves propagation by modulating its phase. When compared to traditional phase or amplitude modulators, this method offers greater benefits as it does not require complex assembly similar to metamaterials and quantum-well structures.[17] It is worth noting that certain organic NLO crystals exhibit piezo resonance peaks at low frequency (below 1MHz) due to converse piezoelectric effect. This internal electromechanical response to applied electric field can enhance the electro-optic efficiency and in turn reduces the high driving electric field required for modulation when operated at this frequency. Nonetheless, there are number of methods available for THz modulation, the modulators operating with less field-biasing remains a challenge.[18–20] Considering the drawbacks of other methods such as low speed, the cryogenic temperature for operation, a single crystalline ferroelectric material, $LiNbO_3$ which was operated in piezoelectric resonance condition was reported for THz modulation. At particular frequency typically in kHz or less than 1 MHz, the material will exhibit a strong lattice-phonon coupling which modifies the phase to control the incident THz wave propagation.[17] This strain-mediated strategy allows accurate, low-power, and mechanically controllable phase tuning and is fundamentally different from carrier-based electrical modulations or photoexcited optical modulation mechanisms. Although previous reviews have highlighted significant progress in THz modulation using various materials and external stimuli [11,13,21], a systematic analysis focusing on strain-driven mechanisms remains limited. In particular, (i) a comprehensive review of THz modulation mechanisms based on mechanical deformation and strain engineering in piezoelectric materials has not been reviewed (ii) a comparative discussion of the performance parameters across different external stimulus-based modulation techniques is still lacking, and (iii) perspectives on use of piezoelectric resonance-based strain control for externally tunable THz modulators for communication remain insufficiently explored.

In this review, we first outline the fundamental principles governing THz wave modulation and define the key parameters that determine device efficiency. We then examine modulation strategies based on external stimuli, including electric field biasing, optical pumping, thermal excitation, and mechanical deformation. A comparative discussion of material platforms, such as semiconductor materials and metamaterials, two-dimensional (2D) materials, flexible substrate and organic NLO crystals is presented. Particular emphasis is placed on strain-driven piezoelectric materials operating under resonance condition, a comparatively underexplored approach that is critically examined and systematically reviewed in this work. Finally, we assess current applications of THz modulators in communication, beam steering, imaging, and sensing systems.

## 2. Basic Mechanism for THz modulation:

The modulation of THz wave depends on the carrier concentration in case of semiconductor and the characteristic behavior of active materials in response to the external fields such as electric field, laser illumination, thermal heating, mechanical strain etc. The external fields create an effect on the refractive index of the materials causing the change in dielectric constant.[15]

The material response when optically excited can be given as;

$$\tilde{n} = n + i\beta \quad (1)$$

where $\tilde{n}$ represents the complex refractive index of the material, $n$ is the real part of the refractive index and β the absorption property. According to the Drude model, the electrical response of the material can be expressed as;

$$\varepsilon(\omega) = 1 - \frac{\omega_p^2}{\omega^2 + i\Gamma\omega} \quad (2)$$

where $\varepsilon(\omega)$ is the complex dielectric function, ω is the incident radiation frequency, Γ is the propagation loss, $\omega_p$is the plasma frequency which can be obtained from;

$$\omega_p = \sqrt{\frac{Ne^2}{m_{\text{eff}}\varepsilon_0}} \quad (3)$$

where N, e, $m_{eff}$, and $\varepsilon_0$ are the density of free carriers, electron charges, the effective mass of electrons and vacuum permittivity respectively. The plasma frequency $\omega_p$ lies in the ultraviolet frequency band for most of the materials which act as metallic. Hence THz waves get attenuated while transmitting through the material or even it cannot propagate. Therefore, lowering $\omega_p$ of the material is essential to obtain high THz band through electrical response. Also, the complex dielectric function $\varepsilon(\omega)$ and complex refractive index $\tilde{n}$ is related as $\varepsilon(\omega) = \widetilde{n^2}$.
Research has been accelerated in search of effective THz phase modulators as similar to metamaterials[20], gated graphene metasurface[22–24], and two-dimensional metamaterials.[25] When an electromagnetic wave with an initial phase and equivalent phase constant of $\varphi_0$ and β propagates in material or medium, its phase φ is stated as $\beta l + \varphi_0$. Therefore, change in phase can be caused due to the change in the equivalent phase constant or propagation length.[16,26,27] Changing the transmission distance is a go-to strategy for microwave frequencies, like in the case of switching line phase shifter that basically uses diodes to choose between transmission lines with varying lengths to regulate phase shift, but adding this feature to an existing structural design is highly complicated and also bringing unwanted parasitic effects. Alternatively, modifying the equivalent phase constant is a much more practical approach that works with the propagation of electromagnetic wave through the material or dielectric properties of the medium.[28,29] An example of this is in the field of point-to-point communication where a terahertz beams directivity can be bumped up to increase the distance it can cover.[30] Another application is multi-beam wavefront switching to make target radar systems capable of multi-area detection.[28] Controlling the phase of an electromagnetic wave is a cornerstone in wave-based technologies.

## 3. Critical Parameters of THz Phase Modulation

In THz modulation, phase shift can be induced in a medium when any external perturbation causes change in the refractive index of the materials. The performance of THz phase modulation can be analyzed using parameters such as modulation depth, modulation speed, band width and insertion loss. The modulation depth of the signal is evaluated using the change in $T_{ON}$ and $T_{OFF}$ states of the signal intensity and obtaining a significant value of modulation depth is essential. The modulation depth can be evaluated using the formula,

$$\mathrm{MD}(\%) = \frac{T_{OFF}-T_{ON}}{T_{OFF}} \quad (4)$$

Modulation depth can also be described as

$$\mathrm{MD(dB)} = 10\, log\, \frac{T_{OFF}}{T_{ON}} \quad (5)$$

For communication systems, modulation depth greater than 7dB provides good transmission properties, however values in the range of 4-5 dB are also acceptable.[25] After years of effort, 100% modulation depth has been obtained through incorporating hybrid bulk semiconductors by increasing carrier concentration in 2D structural design.[31,32] Both the operation principles and the intrinsic characteristics of active materials have a significant impact on modulation speed. Modulation speed is examined by the response of material for fast-varying optical excitations. The speed response is due to the intrinsic properties of material such as electron-phonon interaction and phonon relaxation.[33] Li et al., experimentally demonstrated THz modulation in YAG: Ce over the frequency ranging of 0.2 to 1.8 THz, achieving 4 MHz of modulation speed and modulation depth of 83.8% under pumping laser power of 0.2 W/cm$^2$. The values are then compared with the bare silicon material.[34] Liang et al., demonstrated a monolithically integrated graphene-based modulator producing 100% THz modulation depth for radiation emitted from a quantum cascade laser (QCL) due to efficient THz field - graphene interaction. At selective area, this material exhibits 110 MHz modulation speed compared to the existing terahertz modulators.[35] Nonetheless the reported results are promising, their translation into practical applications like bioimaging and high-speed data transmission remains largely unexplored. Integrating optical modulator and the wave guide can minimize the mismatch of the refractive index, hence it reduces the coupling losses.[25] Rodriguez et.al., reported an insertion loss of about 5% in THz power, corresponding to 0.2 dB per graphene layer in THz transmission. This loss is attributed to minimal optical conductivity of the single-layer graphene in THz regime.[36] Alaloul. et al., proposed plasmon-enhanced graphene modulators which exhibiting the insertion loss of about 6.2 dB keeping a reasonable modulation efficiency of 0.28dB/μm.[37] The integrated graphene-based THz modulator combined with surface emitting concentric-circular-grating QCL exhibited a 3dB electrical modulation bandwidth of about 110 MHz, showing a flat dependency up to 12 MHz.[35]

## 4. External Stimuli-Based THz Phase Modulation

A wide array of materials has been investigated for THz phase modulation, and each of these materials exhibit different physical responses targeting refractive index when exposed to external perturbation. Various attempts using different external stimuli like electric, optical, thermal excitation, mechanical deformation are shown in Table 1. For instance, electric fields affect the molecular orientation and thus refractive index to change the THz output. The vanadium dioxide ($VO_2$) and GST ($Ge_2Sb_2Te_2$) exhibit changes in optical properties while the transformation from amorphous to crystalline phase. For example, the transmittance spectra of GST metasurface in both amorphous (AGST) and crystalline (CGST) phases possesses around 1.75 THz. The modulation index was about 96% indicating its potential as a THz switch at this frequency[38]. Hang Yu and co-workers fabricated a $VO_2$ metamaterial device which can be controlled by temperature. When the temperature was 20°C, the transmittance of the device was above 90% operated at 0.48 THz. While the temperature was increased above the phase transition i.e., 66 °C the absorption rate was more than 99.3% at 1.247 THz at 80 °C showing a perfect absorption effect[39]. These transitions are triggered thermally or optically, making them a promising material for reconfigurable THz optics.[40] The reversibility of the material shows repeatable phase control, even though thermal inertia and heat dissipation can limit switching speed and stability. In recent days, semiconductor heterostructures and 2D materials have gained more attraction. The change in carrier concentration can drastically alter the materials conductivity and dielectric function in the THz regime. Other 2D materials including graphene layer offer ultrafast phase modulation within sub-nanosecond response times. These are at ease and compatible with fabrication processes, integrating them into THz circuits.[35]

### 4.1 Modulation by Electric field

Ren. et al., demonstrated electro-optic THz modulator using electrically driven current-induced transition from insulator to metal in a thin layer of $VO_2$. When milliampere DC current applied between Ag electrodes with size of 75 nm $VO_2$ film, the change of conductivity and impedance has been observed. This current control and allows tuning of transmission, reflection, and absorption, at 0.55 THz. About 180° phase switching under 18 mA external current is observed as shown in Figure 1. The reflection modulation depth of about 99.9% was achieved at low THz frequencies, exhibiting repeatable, reversible response time of 1s and showing stable operation over 100 switching cycles. A feedback “THz-electro-THz” loop is implemented where the reflected THz signal is monitored and automatically adjust drive current to achieve (PR≈0) under different currents and external conditions.[5] Sahoo.et al., reported electrically tunable THz phase modulator based on twisted-nematic (TN) liquid crystal cells. The phase modulation was achieved through molecular reorientation under applied AC voltage across pristine poly (3,4-ethylenedioxythiophene): poly(styrenesulfonate) (PEDOT: PSS) thin film as transparent conductive electrodes (TCEs). The device exhibited ~ 92% transmittance over 0.2-1.2 THz at ~1 μm thickness, which is higher than the doped PEDOT: PSS or nanostructured electrodes. A phase shift of about 95° at 1 THz with sub-10 V drive and low insertion loss was obtained.[41] Liu. et al., demonstrated an ultrawideband solid-state THz phase shifter where the phase is tuned electrically via a graphene conductive interface embedded in total internal reflection (TIR) geometry integrated with a non-resonant metasurface of micro slits. Upon sweeping the voltage between -5 and +5 V, the device achieved an average phase shift of about 120° for 0.4-1.2 THz and over 0.5 THz, the phase shift was found to be ~ 106°.[42]

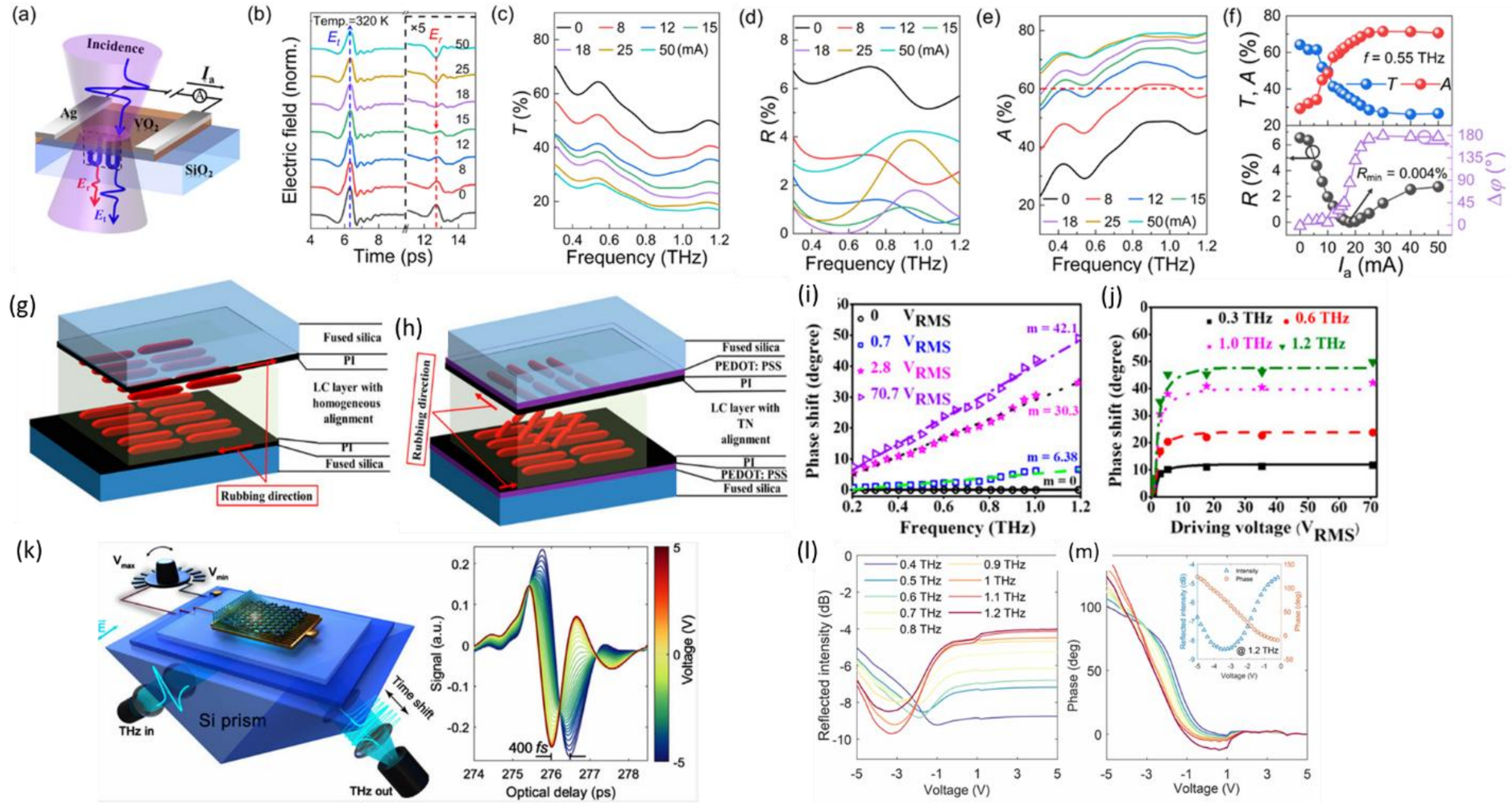

***Figure 1.*** *(a) Schematic of the electrically driven VO₂ THz modulation device. (b) THz time-domain signals of a 75-nm VO₂ film under applied currents (0–50 mA), showing monotonic decrease in transmission and amplitude reversal in reflection. Corresponding THz (c) transmission, (d) reflection, and (e) absorption spectra (f) Transmission, reflection, absorption, and reflection phase shift at 0.55 THz, with a minimum reflection* $R_{min}$ *≈0.004% at 18 mA. Reproduced from reference 5 with permission from American Chemical Society, Copyright 2022. (g) device structure of a homogeneously aligned liquid crystal (LC) cell to carry out the optical constant measurement of both types of LC and (h) a twisted nematic (TN)-LC phase shifter with pristine poly(3,4-ethylenedioxythiophene): polystyrenesulfonate) (PEDOT: PSS) as transparent electrodes. (i) Phase shift properties as a function of frequency for several values of applied voltage and (j) applied voltage at selected frequencies for an MDA-250 TN-LC-based THz phase shifter. Reproduced from reference 41 with permission from MDPI, Copyright 2019. (k) Diagram of the graphene-loaded metal microslits device in the total internal reflection geometry (l) The insertion loss and phase shift of D3 under finer gate voltages. Reflected intensity as a function of gate voltage at different frequencies. The reflected intensity is referred to as the Si prism without the device. (m) Relative phase shift referred to the phase at 5 V from 0.4 to 1.2 THz. The inset shows the linearity of the phase shift as a function of driving voltage and the intensity fluctuation. Reproduced from reference 42 with permission from American Chemical Society, Copyright 2024.*

### 4.2 Modulation by Optical pumping

Terahertz optical modulation is achieved based on optical-excitation-control-terahertz radiation. A continuous wave or pulsed laser of single photon carrying energy is used to excite the active material. Phase modulation can be achieved when photoexcited carriers in doped semiconductors and phase-change metasurface (PSM) structures for e.g., GaAs, doped Si, $VO_2$ composites, alter the materials conductivity and resonance properties. These changes can induce THz phase shift more than 100°. The conductivity change shifts the resonance frequency and mode type of the metasurface unit (e.g., split-ring resonator (SRR) to closed ring resonator), producing a controllable delay in phase transmission or reflection.[16] Xu. et al., demonstrated THz phase modulation using $NdGaO_3$ (NGO) single crystals by tuning their refractive index with temperature and optical pumping as shown in Figure 2. The temperature is varied between 100-400 K, and optical pumping uses a white-light picosecond laser (450-700nm, ~25 ps, ~78 MHz) with fluences up to ~20 J/cm² to focus on the crystal. Among typical oxide substrates, only NGO exhibits a tunable phase shift under temperature and optical pumping, making a bulk single crystal as an efficient THz phase modulator without any metamaterial structuring. It was found that at ~500 µm NGO crystal, 94° phase shift at 1.5 THz was observed as the temperature rises from 100 to 400 K. Under the optical illumination with fluence up to ~20 J/cm² at fixed temperature, a decrease in transmitted THz amplitude about 30% and a temporal delay of ~0.16 ps, which corresponds to a phase shift of ~78° at 1.5 THz was observed. The complex refractive index analysis shows that temperature primarily induces a uniform increase in the refractive index (n) with negligible changes in the extinction coefficient (k), consistent with low-loss thermal phase modulation. Conversely, optical pumping modifies both n and the frequency-dependent k, reflecting the influence of carrier excitation and optical transitions.[18]

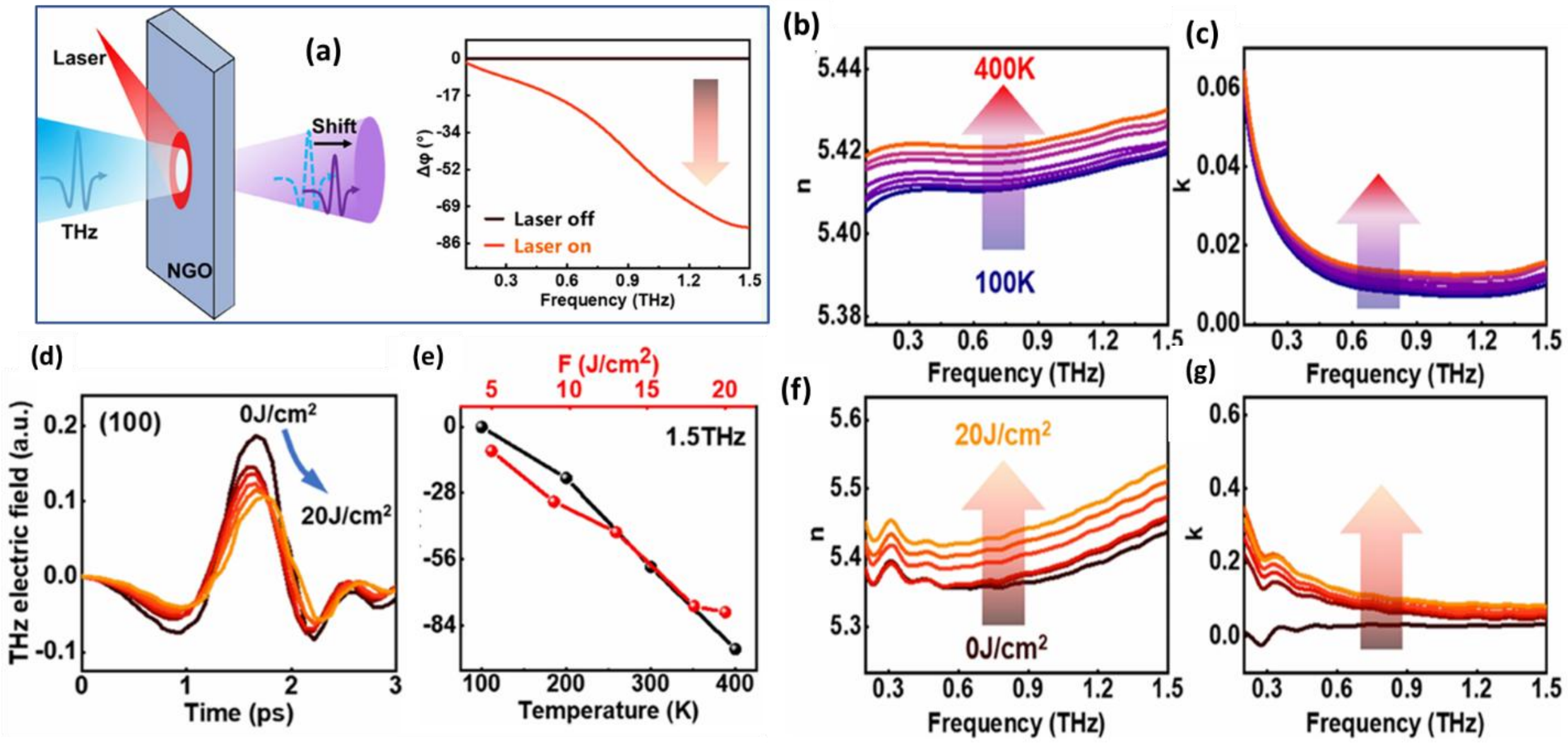

*__Figure 2__. (a) Terahertz phase shift in $NdGaO_3$ single crystal. (b, c) Refractive index n and k measured as a function of temperature. (d) THz time-domain signal measured with different illuminations (0, 5, 9, 13, 18, and 20 $J/cm^2$). (e)Phase shift (Δφ) measured at 1.5 THz as a function of both light fluence (F) and temperature. (f, g) Refractive index n and k measured as a function of light fluence. Reproduced from reference 18 with permission from American Chemical Society, Copyright 2022.*

## 4.3 Modulation by Thermal Excitation

Achieving phase shift by thermal excitation induces change in refractive index, conductivity or structural phase of the active material in response to temperature. Dong. et al., obtained phase modulation both for transmitted and reflected THz waves around 2.5 THz by thermally driving the insulator-metal phase transition of $VO_2$ due to change in conductivity as shown in Figure 3. When temperature rises from 340-350 K, the $VO_2$ shows transition from low-conductivity dielectric to high conductivity metal, thereby transforming the structure from transmitting FSS into highly absorbing, reflective metamaterial imposing a controllable and stable phase shift on the propagating field.[43]

Generally, a standard derivation of optical phase through a slab is given as;

$$\boldsymbol{\phi} = \frac{2\pi n d}{\lambda} \tag{6}$$

Considering the change in the refractive index Δn, caused by a temperature change ΔT and material possessing thermo-optic coefficient, then $\Delta n \approx \frac{dn}{dT}\,\Delta T$ (for small $\Delta T$)[44],

$$\Delta\phi = \frac{2\pi}{\lambda}\left(\frac{dn}{dT}\right) d\,\Delta T \tag{7}$$

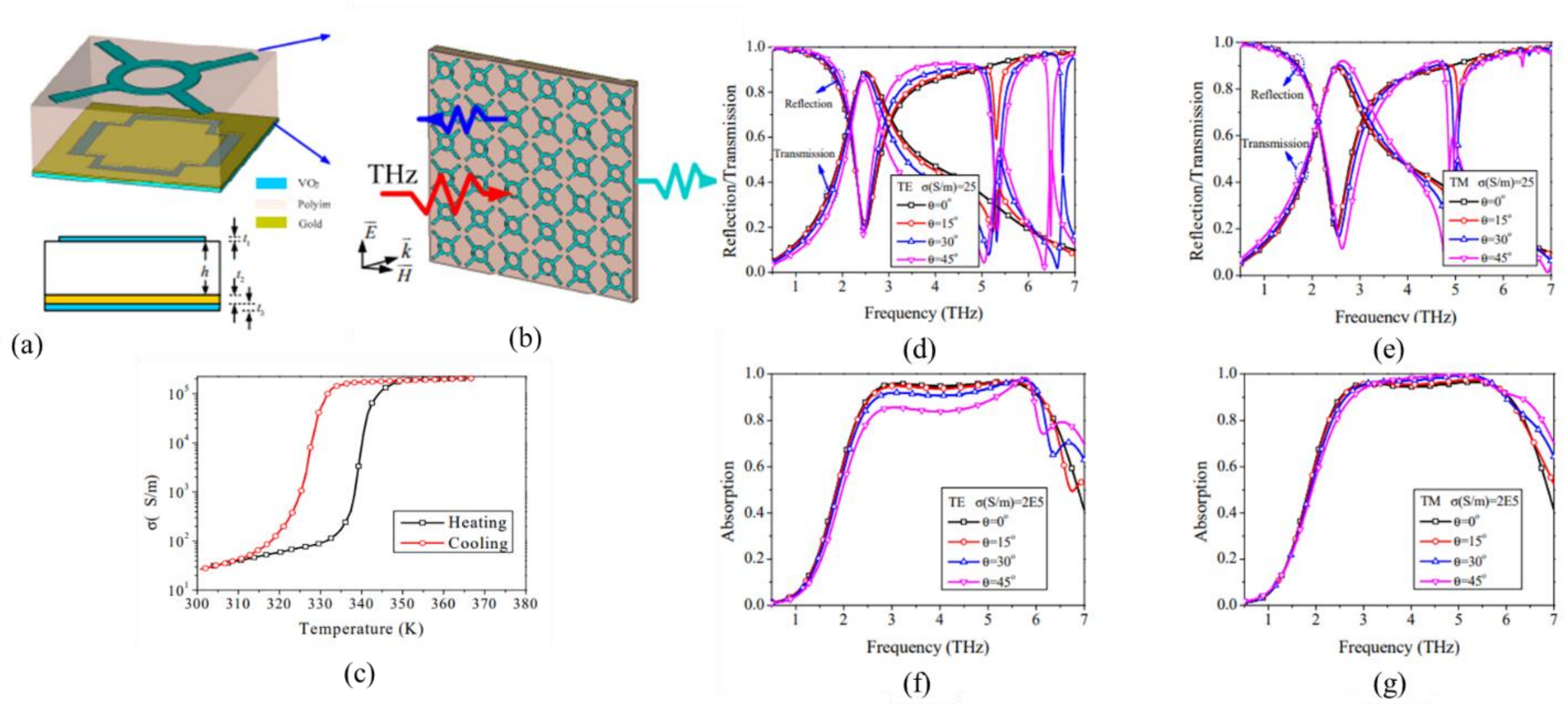

*__Figure 3.__ (a) 3D schematic of the proposed unit-cell design and (b) corresponding 2D array layout. (c) Measured conductivity of $VO_2$ at different temperatures. (d, f) Simulated transmission, reflection, and*

*absorption responses for various incident angles under TE (e, g) and TM polarizations for different $VO_2$ conductivity states. Reproduced from reference 43 with permission from MDPI, Copyright 2021.*

## 4.4 Modulation by Mechanical Deformation

Phase modulation by mechanical deformation mainly relays on the THz metasurfaces whose resonance phase is tuned by reconfiguring their geometry by implementing MEMS-like actuation, without altering their optical properties. The key parameters of metamaterial unit cells like gaps, overlap, angle, height, and separation modify the effective capacitance and inductance. Since the resonance frequency and phase response of each unit cell depends on the LC circuit, mechanical deformation such as bending, tilting, or translating enables controlled change of the transmitted/reflected phase at a fixed operating THz frequency. The structural design plays an important role in modulating the input signal to desired output signal intensity. Designs based on cantilever arrays, complementary split-ring/cantilever hybrids structure simultaneously alter the group delay, makes shift in resonance frequency which in turn enables phase and dispersion control around the narrowband THz resonance. Although mechanically driven THz metamaterials exhibit modulation speed on the order of kHz, their performance is fundamentally limited by mechanical response times and fabrication constraints, making them significantly slower than modulation schemes governed by intrinsic electronic dynamics (Figure 4).[21] Tao.et al., demonstrates tunable magnetic and electric resonances around 0.5 THz producing strong phase shift in transmission by varying the out-of-plane angle from ~0° to ~65–80°. Therefore, mechanical deformation approach provides large, continuous tuning of EM response by geometric reorientation, rather than relies on the low carrier-density changes in the materials.[45] Manjappa. et al., achieves phase modulation by mechanically deforming MEMS cantilevers in a Fano metasurface by implementing two bimorph split-ring resonators (SRR-1 and SRR-2). While applying voltages $V_1$ and $V_2$, the SRR gets electrostatically down resulting in the deformation of height $h_1$ and $h_2$. This deformation tunes the coupling between SRR-1 and SRR-2, initiating between a broad dipole resonance and a sharp Fano resonance causing shift in their resonance frequencies (e.g., Fano around 0.56–0.58 THz and dipole around 0.67–0.77 THz). However, when it is operated beyond the pull-in voltage (> ~25 V), the cantilevers get stick to the substrate that breaks repeatability and slows reconfigurability. [46]

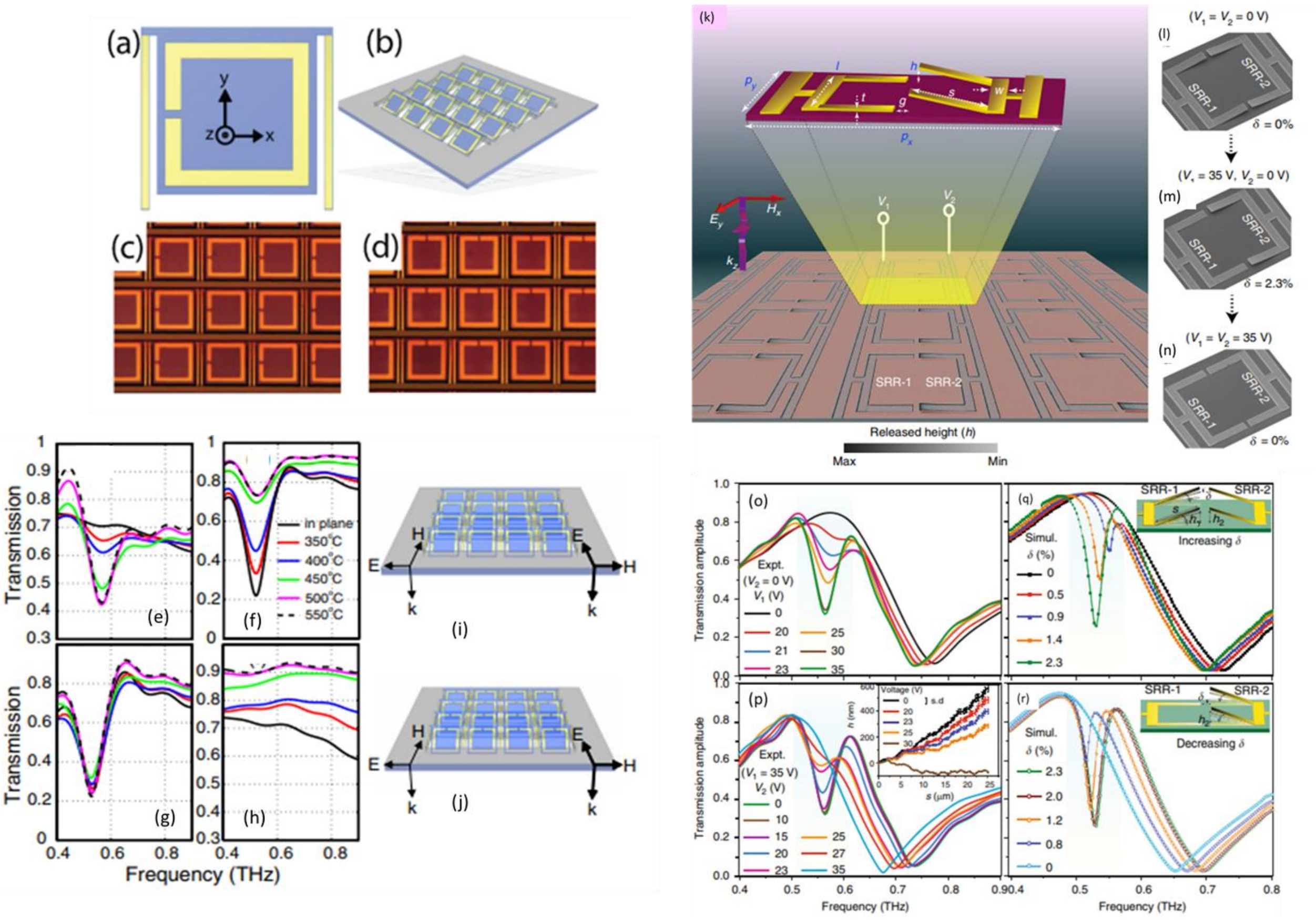

***Figure 4.*** *(a) Unit cell consisting of a split ring resonator and cantilever legs. (b) Schematic view of a portion of the metamaterial structure highlighting how the SRRs rotate as the cantilever legs bend. (c) and (d) Photographs of portions of two of the THz metamaterial arrays fabricated for this study. The SRRs are 72 µm × 72 µm with an in-plane periodicity of 100 µm and an overall dimension of 1cm × 1cm. (e-j) Magnetic Response: Transmission as a function of frequency for various orientations of the SRRs. The black curve is with the SRRs lying in the plane of the substrate, red curve—RTA 350° C, blue curve—RTA 400° C, green curve—RTA 450° C, magenta curve—RTA 500° C, dashed black curve—RTA 550° C. (f) Electric Response: color code is same as for (e). (g) Schematic showing the polarization for (e) (on left) and (f) (on right). (h) Bianisotropic response: color code as for (e) and (f). (i) Nonresonant response. (j) Schematic showing the polarization for (h) (on left) and (j) (on right). Reproduced from reference 45 with permission from American Physical Society, Copyright 2009. (k) Coloured scanning electron microscope (SEM) image of the MEMS Fanometasurface. The unit cell comprises of two SRRs separated by a gap g, where SRR cantilever arms of length s are released at a height h. The unit cell dimensions are depicted in the inset, where $p_x$: 110 µm; $p_y$: 75 µm; l: 60 µm; s: 25 µm; w: 6 µm; g: 4 µm; and t: 900 nm. V1 and V2 are the input voltage ports to achieve the independent actuation of SRR-1 and SRR-2, respectively. l–n SEM images of the unit cell showing the sequential actuation of SRRs with voltage V1 and V2 applied across the two SRRs, where the sequence from (l) to (m) represents the increasing asymmetry (δ) and (m) to (n) represents the decreasing asymmetry configuration. (o-r) Active tuning of Fano resonances in MEMS metasurface. Reproduced from reference 46 with permission from Springer Nature, Copyright 2018.*

***Table 1:*** *Comparison of the reported $VO_2$-based active THz modulators*

| Sample type | Thickness of layer (nm) | Modulation type | Trigger type | Maximum modulation depth (%) | Qualified bandwidth (THz) | Triggering threshold | Modulation speed | Ref. |
|---|---|---|---|---|---|---|---|---|
| $VO_2$ | 190 | transmission | electrical and optical | 80 | 0.2 − 2 | 30 V | 26 ms | [47] |
| $VO_2$/ CNT | N/A | transmission | electrical and optical | 91 | 0.2 − 2.5 | 0.58mW/ $mm^2$ | 27 ms | [48] |
| $VO_2$ | 280 | reflection | thermal | 99.9 | 0.3 – 0.9 | 338 K | N/A | [49] |
| $VO_2$ | 100 | reflection | optical | 96.7 | 0.2 −1 | 5.6mJ/$cm^2$ | N/A | [50] |
| $VO_2$-metasurface | N/A | transmission | electrical | 54 | 0.55 −0.7 | 1 A | 1.5 s | [51] |
| $VO_2$/PVP | N/A | transmission | optical | 65 | 0.32 −0.5 | 4.3mW/$cm^2$ | N/A | [52] |
| $VO_2$ | 192 | transmission | optical | 81.2 | N/A | 2mJ/$cm^2$ | 120 ms | [53] |
| $VO_2$/ graphene | 290 | reflection | thermal | 96 | 0.3 −1 | 300 K | N/A | [54] |
| W: $VO_2$ | 50 | transmission | thermal | 80 | 0.3 −2.3 | 353 K | N/A | [55] |
| $VO_2$ | 75 | reflection | electrical | 99.9 | 0.35 −0.76 | 18 mA | 1 s | [5] |

# 5. Material design for THz modulation

## 5.1 Modulator Based on Semiconductor Materials and Metamaterials.

The electromagnetic waves can be artificially manipulated by designing the structure and controlling the size of unit cells in semiconductor materials and metamaterials. Tanaka et al. used 2D meta hole arrays (2D-MHAs) as a THz modulator in 2004 which opened up the way for many research groups to utilize the laser-based approach for controlling THz wave.[56] Drude model can be used to explain the mechanism of semiconductor materials under the excitation of electric field where the free electrons will be treated as free electron gas in THz wave region. At room temperature, free carriers are generated when photon energy ($h\omega$) exceeds the bandgap of semiconductor materials, leading to strong attenuation of THz wave or microwaves. In contrast, the materials become transparent when the operating frequency is greater than the plasma frequency $\omega \geq \omega_p$ or at ultraviolet region. Therefore, reducing the plasma frequency is important when transparency is required at lower frequencies such as the THz or microwave region. The change in effective mass of carriers and density of the electron can also tune the characteristic frequency. Therefore, a laser with energy higher than a material's bandgap can be used to modify

THz waves. Effective modulation can be attained by integrating the semiconducting material with various structural designs, like photonic crystal, surface plasma array and anisotropic medium. Liet.al., reported a high-resistance Si wafer as modulator. [57] The modulation speed of 0.2 kHz was only achieved, even though the lifetime of carriers for Si wafer is long for material designed with ultra- high resistivity. In contrast, the lifetime of carriers in GaAs is short. Fekete et al. reported a modulator using stacked $SiO_2$ photonic crystal with embedded GaAs defect layer.[58] Here, an ultrashort 810 nm laser pulse can be used to excite the front GaAs surface at low concentration, resulting in an effective modulation of the THz beam. In Sb gratings on a semi-insulating GaAs substrate, Deng et al. showed a modulation speed of 1.2 GHz (Figure 5) (a, b).[59]

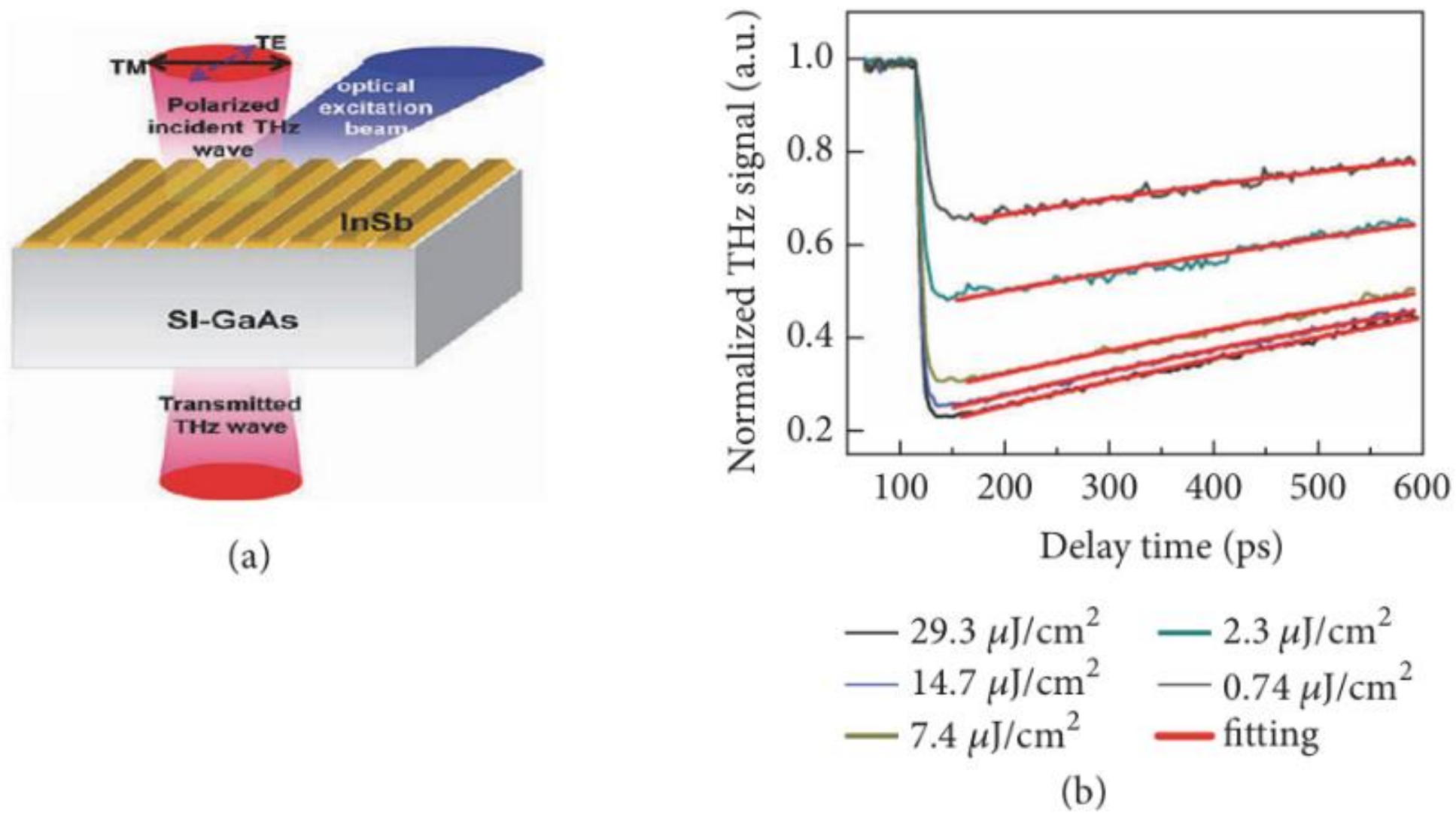

***Figure 5**. Optically tuned THz modulator based on semiconductor materials and metamaterials. Panel (1). (a) SEM images of the fabricated InSb grating. (b) Normalized THz signals as a function of delay time measured by OPTP under different pump laser fluences and the exponential curve fittings. Reproduced from reference 59 with permission from John Wiley and Sons, Copyright 2013.*

### 5.2 Modulator Based on 2D Materials

Two-dimensional (2D) materials such as graphene exhibit a unique electrical and optical properties, which have gained significant interest for use in optoelectronic applications.[60] The arrangement of carbon atoms gives graphene its characteristic conical band structure. Graphene's conductivity is prominently contributed by the in-band and band-to-band transition of electrons. In-band transition of electrons dominates in THz range, due to small photon energy. Consequently, the Drude model can be used to estimate graphene's thin layer conductivity as [61]

$$\sigma(\omega) = -\frac{jD}{\pi(\omega - i\Gamma)} \tag{8}$$

Therefore, the Fermi level and graphene's conductivity are connected. In general, the concentration and kind of carriers can be dynamically changed by varying the Fermi level. The Fermi level shifts towards the conduction band due to the increase in electron concentration and shifts towards the valence band due to the increase in hole concentration. A low conductivity of graphene will be observed when the Fermi level lies at the Dirac point, as the electron and hole densities reach equal. The first demonstration for optical modulation was performed by Zhang Xiang's et.al., by regulating the Fermi level of graphene.[62] A notable study was observed using a light-controlled graphene devices based on graphene deposited on silicon (GOS) (Figure 6). Under the excitation of 750 nm with power of 40 mW, the modulator could achieve a modulation depth of about 68% for a tunable THz bandwidth ranging from 0.2-2 THz. The transmission of THz wave nearly vanished when the optical pump energy is increased to 500 Mw (@750 nm). The graphene layer absorbed approximately 2.3% of the modulated optical beam.[63]

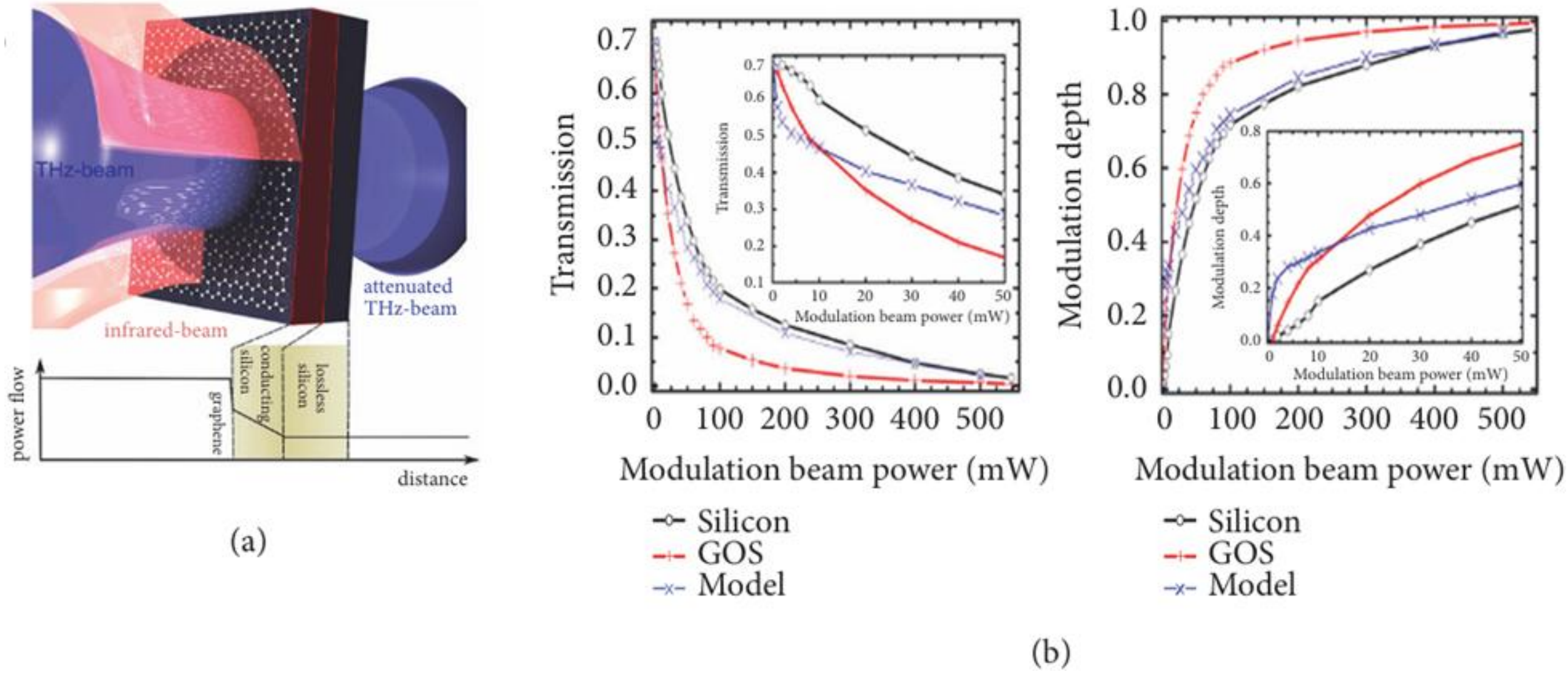

*Figure 6. Optically tuned THz modulator based on 2D materials. (a) Schematic view of graphene on Si sample. (b) Normalized transmission (left) and depth (right) from the three phthalocyanine structures. Reproduced from reference 63 with permission from American Chemical Society, Copyright 2012.*

Considering the stacking of 2D crystals, mismatch in the lattice constants and misalignment of crystallographic axis leads to a formation of moiré pattern. This moiré effect is observed in twisted bilayer graphene (TBG) due to the small mismatch between the adjacent graphene layer which exhibits strongly correlated electronic states, tunable flat bands, and enhanced light–matter interactions. An earlier experimental demonstration by Xingquan Zou and co-workers reported the conductivity peak near 2.7 THz on the conventional Drude response of graphene. This was attributed to the enhanced density of states by Hove singularities originated from the TBG [64]. An optically pumped THz modulator utilizing Si-grown MoS2 metasurface was described by Zheng et al. The obtained modulation depth of MoS2 metasurface reaches over 90% utilizing a CW laser pumping of 4W/cm2 power density, where the modulation efficiency is better than without $MoS_2$ [65]. Integrating few thin layers of graphene in $WSe_2$ monolayer with suitable design of van der waals heterostructures stack produced an efficient THz phonon generation of upto 3 THz [66]. Such heterostructures are promising for tunable optoelectronic interface offering a scalable platform for integrating 2D materials into next-generation THz and infrared photonic systems [67].

### 5.3 Modulator Based on Flexible Substrate.

THz modulators based on flexible substrates provide several advantages including transparency, lightweight structure, low cost, and good adhesion, making them potential for THz modulation.[68] D. Y. Khan et. al., fabricated a single crystal Si with a thickness ranging from 20 nm to 320 nm into a Si ribbon consisting width of 5 $\mu$m to 50 $\mu$m and a length of 15 mm (Figure 7).[69] When the structure is integrated with a flexible substrate, this wavy Si ribbon can undergo reversible stretching and compression while maintaining its structural integrity even at high horizontal stress. The electrical parameters of Si ribbon can be controlled by changing the surface configuration of the flexible substrate. Kebin. et.al., proposed an optically controlled metamaterial modulator fabricated on a flexible polymer sheet that can be operated at THz frequency.[70] The modulator is designed by integrating electric split-ring resonators (eSRRs) on a thin polyimide layer. When GaAs patch was optically excited, the response of the metamaterial was varied due to the change in dielectric constant. In the frequency range of 1.1-1.8 THz, a 60% modulation depth was attained. These flexible devices can be implemented on nonpolar structures, enabling their use in variety of applications. Table 2. provided a comparative overview of various material systems and modulation strategies employed for THz modulation, highlighting their modulation depth, response speed, operational bandwidth, and driving conditions.

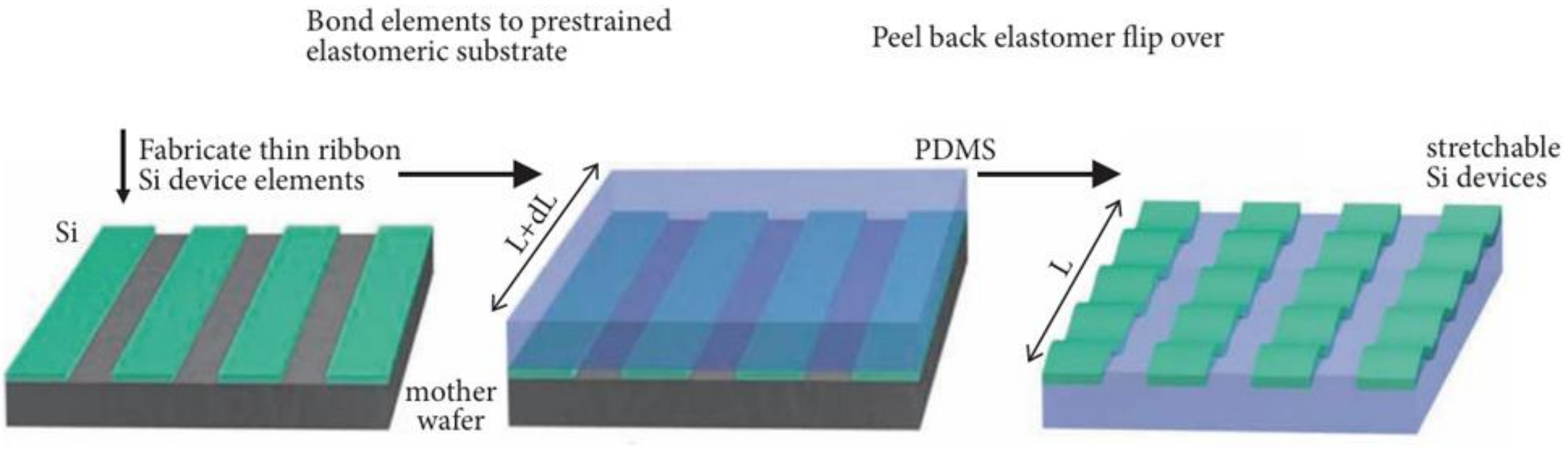

***Figure** 7. Optically tuned THz modulator based on flexible substrate. Schematic illustration of the process for building stretchable single-crystal Si devices on elastomeric substrate. Reproduced from reference 69 with permission from The American Association for the Advancement of Science, Copyright 2006*

.

***Table 2:*** *Comparison of various material structures in THz modulation*

| Material and structure | Operating principle | Modulation depth (%) | Modulation speed or response time | Modulation bandwidth | Operating condition |
|---|---|---|---|---|---|
| Graphene–Si | A-O | 69 | — | 0.2–2.0 THz | 40 mW |
| Graphene–Si | O-E | 83 | ~1 kHz | — | 420 mW / −4 V |
| Graphene–Si | O-E | 93 | ~4.6 kHz | — | 100 mW / 10 V |
| Graphene–Ge | A-O | 94 | ~200 kHz | 0.025–1.0 THz | 800 mW |
| Graphene | M | 26 | — | — | 10^-2 GPa |
| Graphene | E | 15 | 20 kHz | — | 0–50 V |
| Graphene | E | 64 | ~4 kHz | — | −10 to 20 V |
| Graphene | E | >99.3 | — | 0.5–1.6 THz | −12 to 16 V |
| Graphene-based metamaterial | E | >47 | ~100 kHz | — | 350 V |
| Graphene-based metamaterial | E | ~50 | ~MHz | — | −20 to 20 V |
| Graphene-based plasmonic | E | 4 | >100 MHz | — | 115 V |
| Graphene-based plasmonic | E | ~100 | — | ~2–12 THz | — |
| Graphene-based optical cavity | E | >95 | — | 0.2–1.6 THz | −60 to 60 V |
| Graphene-based capacitor | E | 99 | — | 0.1–2.5 THz | 3 V |
| Graphene-based waveguide | E | 50 | 2 kHz | — | −35 to 35 V |
| $MoS_2$–Si | A-O | >75 | — | — | 0.24 W/cm² |
| $WS_2$–Si | A-O | 94.8 | <1 ms | 0.2–1.6 THz | 470 mW |
| $WS_2$–Si | A-O | 99 | — | 0.25–2 THz | 2.59 W/cm² |
| $MoTe_2$–Si | A-O | 99.9 | $\tau_{rise}$ = 21 ms; $\tau_{fall}$ = 5.2 ms | 0.3–2.0 THz | 300 mW |
| $PtSe_2$–Si | A-O | 32.7 | — | 0.1–1.0 THz | 1 W/cm² |
| $TaS_2$–Si | A-O | 46.8 | $\tau_{rise}$ = 0.67 ms; $\tau_{fall}$ = 0.1 ms | 0.1–1.0 THz | 1 W/cm² |
| Multiple quantum well | A-O | 40 | — | 0.2–1 THz | 0-2.2 mW |
| InSb gratings | A-O | 46.70 | 1.2 GHz | 1.5 –2.5 THz | 120 mW |
| Graphene on Ge | A-O | 94 (0.1 Db/um) | 1.2 GHz | 0.25–1THz | — |
| $MoS_2$ | A-O | 96 | <1 ps | 0.5–1.5 THz | 4.56 W |
| $WS_2$ | A-O | 99 | — | 0.25–2 THz | 2.59 W/cm$^2$ |
| $C_{60}$ | A-O | 98 | — | 0.5–1.5THz | — |
| AlClPc | A-O | 99 | — | 0–2.5 THz | 1.57 W/cm$^2$ |
| Si Nanotip | A-O | >90 | — | 0.25–1 THz | — |
| Flexible substrates | A-O | 100 | — | 0.2–1.5 THz | — |
| Liquid crystals | T | 90 | — | 1–3 THz | — |

Abbreviations: A-O, all-optical; E, electrical gating; Ge, germanium; M, mechanical strain; O-E, optical-electrical; T, Thermal.

## 6. Strain Driven Piezoelectric Material for THz Modulators

Organic NLO crystals have a significant demand in various applications in THz photonics such as THz generation and detection. Particularly, in non-centrosymmetric crystals, an applied electric field induces mechanical strain in the crystal lattice which attributes to converse piezoelectric effect. At certain frequencies, the crystals natural vibration gets strongly coupled with the applied electric signal resulting in resonance peaks, termed as piezo resonance peaks. These resonance peaks are observed in the low frequency region (below 1 MHz) showing sharp peaks at impedance spectrum or dielectric analysis. These resonance peaks also depend on the thickness of the crystal. This strain can significantly alter the refractive index of the material through electro-optic effects. Hence driving the piezoelectric crystal at the resonance condition enhances the electro-optic coefficients enabling efficient modulation of THz waves through induced polarization and strain. Organic piezoelectric crystals such as 4-N, N dimethylamino-4'-N'-methyl-stilbazoliumtoyslate (DAST) [71], 4-N, N-dimethylamino-4-N-methyl-stilbazolium 2,4,6 trimethylbenzenesulfonate (DSTMS) [2], 2-(3-(4-hydroxystyryl)-5,5-dimethylcyclohex-2-enylidene) malononitrile) (OH1) [72], 2-cyclo-octylamino-5-nitropyridine (COANP) [73] exhibits high electro-optic coefficient. The electro-optic coefficient of DAST is very high, with $r_{111}$ approximately 92±9 pm/V at wavelength of 720 nm and remaining crystals have around 47±8 to 53±6 pm/V near the telecom wavelength (1313–1535 nm) which is significantly larger than inorganic crystals such as ZnTe or $LiNbO_3$. Owing to their high nonlinear optical properties, DAST produces about 0.1–8 THz bandwidth with d33 ~ 120 pm/V while DSTMS and OH1 shows a spectral range of 0.1– 4 THz and 0.1–3THz with the $r_{11} \approx$ 37 pm/V and $r_{33} \approx 109 - 52$ pm/V respectively. The nonlinear-optical susceptibility of $d_{33}$ = 13.7± 2pm/V were reported for COANP crystal. All these crystals remain advantageous with the conventional inorganic crystals in terms of ultra-broadband THz generation.

Shankar et.al., reported that piezoelectric (-)-2-α-methylbenzylamino-5-nitropyridine (MBA-NP) crystals exhibits piezoelectric vibrational frequencies that strongly depend on the sample thickness. The frequency response of the dielectric constant ($\varepsilon_r$), dielectric loss tangent (tanδ) and impedance (Z) were measured for two different sizes of sample resulting in different range of resonance frequencies around 100-200 kHz and 70-90 kHz as shown in Figure 8. When an external field is applied, coupling of stationary waves and applied electric field leads to the resonance effects. Therefore, these observed resonance peaks attribute to the converse piezoelectric effect in the crystal at its resonant frequencies. A noticeable shift in the resonance peaks towards the lower frequencies with increase in temperature was observed. The increase and shift in resonance frequencies as the temperature increases due to the converse piezoelectric effect has exploited the crystal for resonant-piezoelectro-optic light modulator applications.[74]

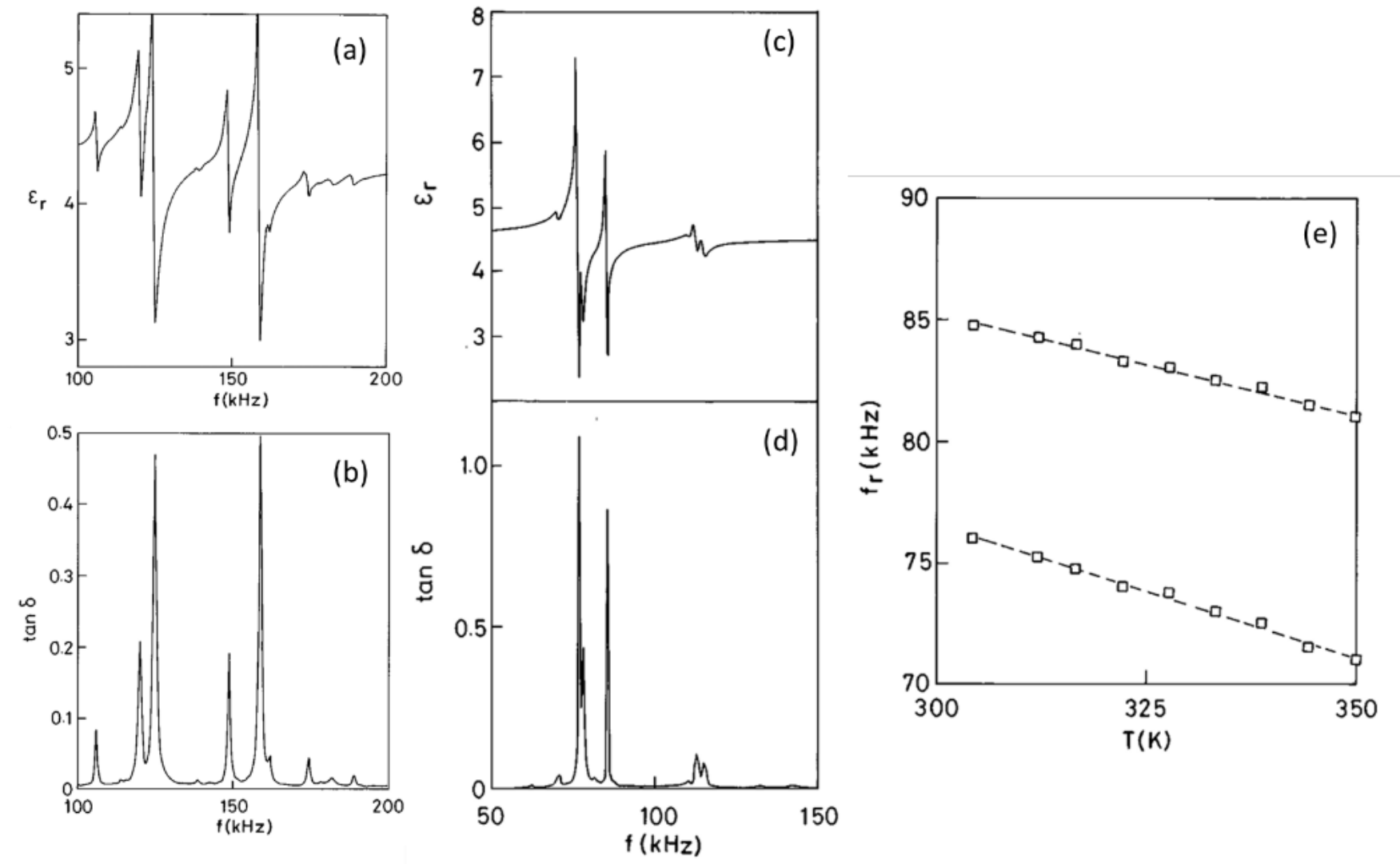

***Figure 8***. *Dielectric response of (-)-2-α-methylbenzylamino-5-nitropyridine crystal with different dimensions, sample 1 of 7.87× 7.39×4 mm$^3$ and sample 2 of 12.38×10.4×2.61 mm$^3$. (a)and (b) shows the frequency response of $\varepsilon_r$ and tan δ along [010] for sample 1. (c) and (d) shows the frequency response of $\varepsilon_r$ and tan δ along [010] for sample 2. (e) shows the temperature dependence of the resonance frequencies. Reproduced from reference 74 with permission from AIP Publishing, Copyright 1997.*

Our research group has already reported similar piezo resonance peaks in the single crystal of 2,6-diaminopyridinium–4-nitrophenolate–4-nitrophenol. Resonance peaks at low frequency were observed in dielectric loss and dielectric constant for different crystal thickness as shown in Figure 9 (a, b). There was an observable shift in the resonance peaks towards the lower frequency i.e., from 9.8496 kHz to 6.4893 kHz as the temperature increases from 303 K to 333 K. Thus, these resonance peaks at lower frequencies have a significant effect on electro-optic measurement. Therefore, operating the device at the piezoelectric resonance frequency can lower the high driving electric field needed for electro-optic modulation. [75,76] Similarly, decrease in resonance frequency when temperature raised was observed in glucuronic acid γ-lactone crystal as shown in Figure 9(c).[77] Saripalli et.al., has assumed the sample as damped harmonic oscillator. A large-amplitude vibration of the sample can be seen when the applied electric field oscillates at a frequency that is near to the sample's natural frequency. Due to the large sequence of compression and refraction in the sample, different longitudinal modes are formed. The frequency of the longitudinal modes is expressed as;

$$f_n = n\frac{v}{2t} \tag{9}$$

where $v$ is the speed of sound in the sample, t is the thickness of the sample, and n takes positive integer values. When the frequency of the applied electric field resonates with the frequency of these modes, piezo resonance is obtained. Therefore, the resulting resonance frequencies are strongly dependent on the thickness and elastic coefficient of the sample. The elastic coefficient is inversely proportional to the square of resonance frequency. The observed variation in the resonance frequency with temperature can be used for sensor applications. A semi organic crystal of sodium p-nitrophenolate dihydrate was reported for similar dielectric behavior where resonance peaks at lower frequency observed at different planes [100] and [010] as shown in Figure 9(d, e).[78] Kerkoc et.al., reported high piezoelectric coefficient of about $d_{13} = 2.2 \pm 0.1$ pm/V and $d_{33} = 8.2 \pm 0.4$ pm/V for 2-furyl methacrylic anhydride.[79]

Raghavendra et. al., carried out dielectric measurement for a metal organic single crystal lithium L-ascorbate dihydrate of thickness 1.94 mm at [010] plate with silver coated surface [80]. The resonance frequency was observed in the region of 100-200 kHz, peaking at 137 kHz as shown in Figure 9 (f). A similar behavior was also reported by Ranjith et.al., for organo-metallic single crystal trans-Diaquabis (glycine) copper (II) Picrate Hydrate (tDgcPH) [81].The dielectric constant value plotted with log frequency for tDgcPH crystal is shown in Figure 9 (g).

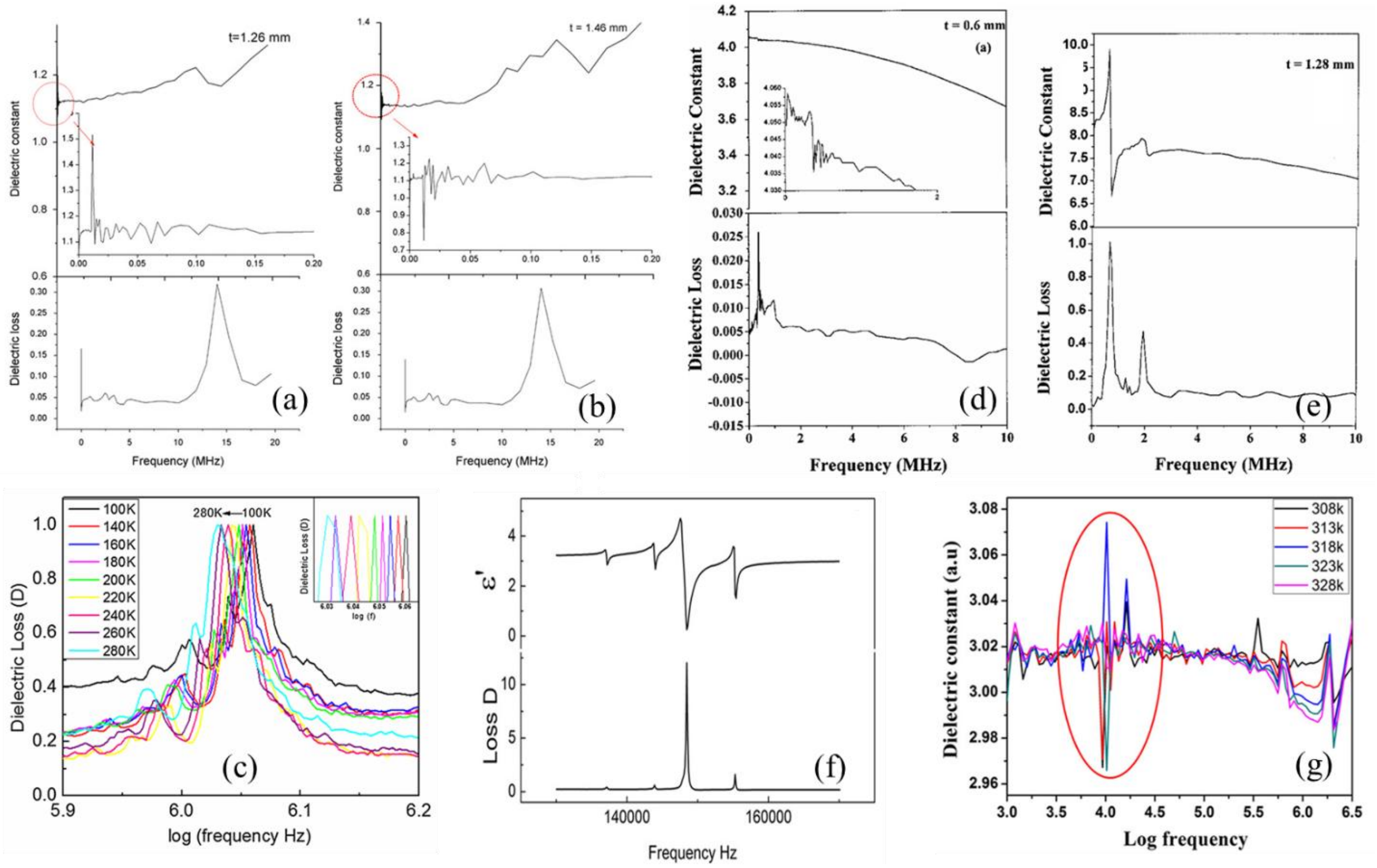

*__Figure 9__. Dielectric measurements for various electro-optic single crystals. (a) and (b) shows the dielectric constant and loss for 2,6-diaminopyridinium–4-nitrophenolate–4-nitrophenol single crystal for different thickness of t = 1.26 mm and 1.46 mm. Reproduced from reference 75 with permission from Elsevier, Copyright 2012. (c) dielectric loss with log frequency for glucuronic acid γ-lactone crystal. Reproduced from reference 77 with permission from Springer Nature, Copyright 2016 (d) and (e) shows the dielectric constant and loss for sodium p-nitrophenolate dihydrate crystal for different thickness of t = 0.6 mm and 1.28 mm. Reproduced from reference 78 with permission from AIP Publishing, Copyright 2000 (f) Piezoelectric resonance peak observed in dielectric dispersion for [010] plane in lithium L-ascorbate dihydrate crystal. Reproduced from reference 80 with permission from Elsevier, Copyright 2013. (g)*

*dielectric constant versus log frequency for trans-Diaquabis (glycine) copper (II) Picrate Hydrate single crystal. Reproduced from reference 81 with permission from Elsevier, Copyright 2016.*

When an electric field, $E_k$, is applied to the crystal, the refractive index changes according to

$$\Delta\left(\frac{1}{n^2}\right)_{ij} = \sum r_{ijk}E_k \tag{10}$$

where, $r_{ijk}$ is the electro-optic tensor. This tensor also depends on the crystal symmetry and is closely related to the piezoelectric tensor [82].

Considering the converse piezoelectric effect, the strain tensor can be given as $S_{ij} = d_{ijk} E_i$. where, the piezoelectric coefficient ($d_{ijk}$) is proportional to second order nonlinearity component ($\chi^2$) [17].

In the THz range, the effective nonlinear coefficient is constant and can be calculated via

$$d_{jkl} = -\frac{n^4}{4r_{jkl}} \tag{11}$$

where n is the refractive index and $r_{jlk}$ the linear electro-optical coefficient. If both two waves in the direction of z mix with each other, the polarization can be calculated as [83] $P_z = 2\varepsilon_0 d_{33} E_z^2$

The nonlinear polarization of the medium, which is connected to the field strength of the excitation pulse, determines the field strength of the THz pulse that is emitted.

$$E_{THz} \propto P_{NL}(t) = \sum_{j,k} \varepsilon_0 \chi^{(2)}_{ijk} E_j(t) E_k(t) \tag{12}$$

Therefore, the converse piezoelectric effect and terahertz emission are linked through strain-induced refractive index changes, $S_{ij} = d_{ijk} E_i$ (where, dijk $\propto \chi^{(2)}_{ijk}$)

The strain modulates the optical path length ($\Delta$L) of terahertz waves because the

$$\Delta\phi = \frac{2\pi}{\lambda}\Delta L \tag{13}$$

where $S_{ij} = \frac{\Delta L}{L}$, and then the $\Delta L = LS_{ij}$

$$\Delta\phi = \frac{2\pi}{\lambda} L d_{ijk} E_i \tag{14}$$

where $\Delta\phi$ is the phase shift and λ is the THz wavelength

Thus, the THz phase modulation directly depends on $d_{ijk}$, the converse piezoelectric effect.

Dutta. et al., has experimentally evidenced the THz phase modulator where a low frequency, strain driven piezoelectric resonance in $LiNbO_3$ (LN) significantly controls the phase of a reflected THz beam ranging from 0.1- 0.5 THz.[17] There is a clear phase shift observed at the piezoresonant node, which explicitly emphasizes the relationship between mechanical displacement and THz phase response. An AC electric field was applied through two surface electrodes. The excitation field of frequency ω was applied in-plane and the out of plane and the measured displacement was observed corresponding to input field. A maximal deflection of 108 pm was found at 20.7 kHz as shown in Figure 10. While a low -frequency AC field (19.2-20.9 kHz) was applied to a multilayer Z-cut $LiNbO_3$ thin film, a maximal out-of-plane surface displacement at 20.7 kHz due to piezoelectric mode was observed as shown in Figure 11. The reflected THz phase shows a strong field in dependence of frequency at a specific band (0.1-0.5 THz) with a maximum response of phase reversal at 0.285 THz. The dependence of THz phase on frequency is strongly correlated with the vibrometric displacement contour, confirming that the strain induced by piezoresonance due to the refractive index govern the THz phase transfer function. To further validate the experimental results, finite element analysis (FEA) simulation using COMSOL Multiphysics suite 4.4 has been utilized. For a range of 19 kHz to 25 kHz, a frequency domain analysis utilizing the physics of piezoelectric devices was executed.

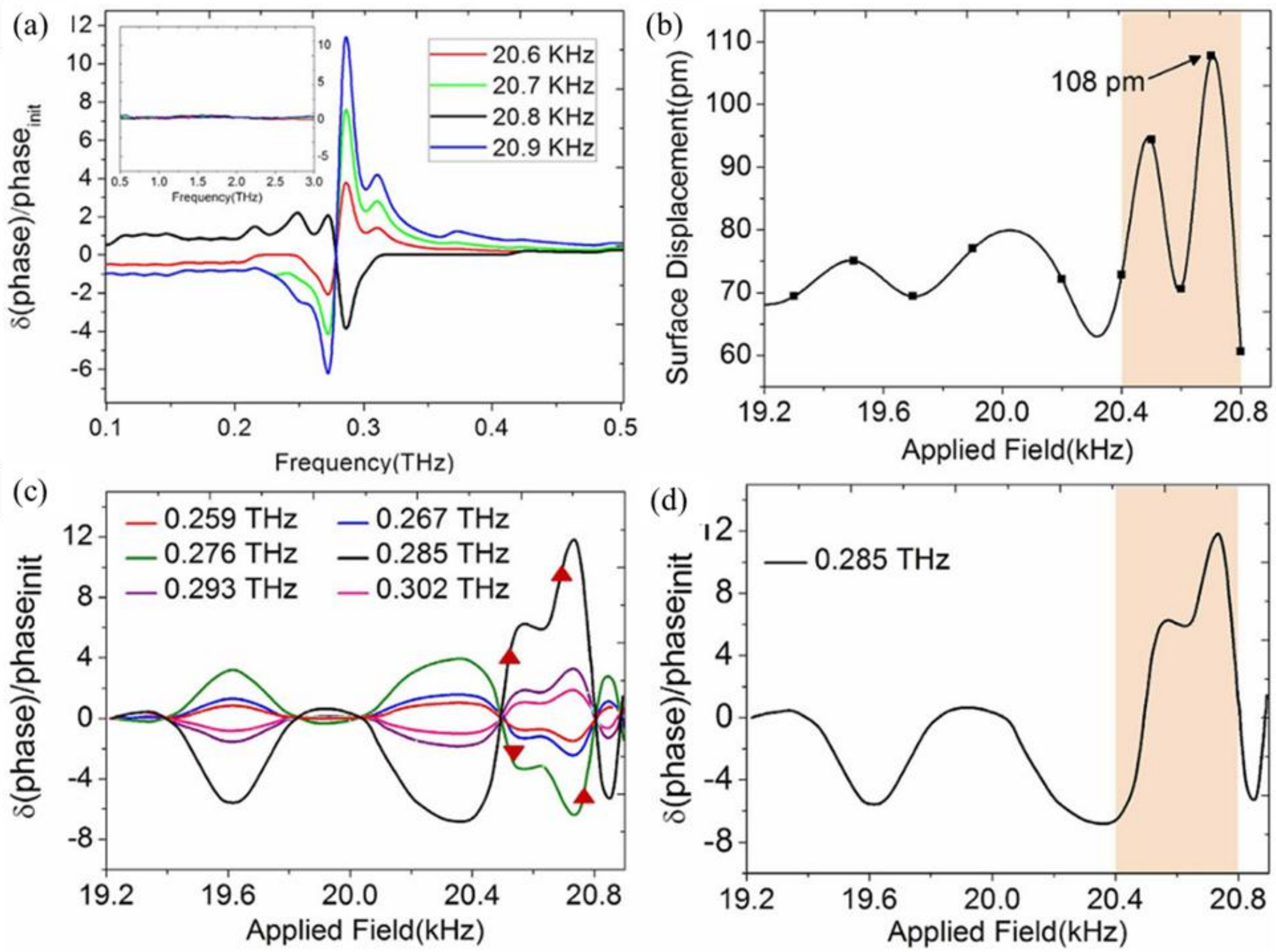

***Figure 10.*** *Evidence of resonance induced phase modulation. (a) Applied field dependent relative phase response of the reflected THz wave, normalized with respect to the response observed without applying any external field. Phase response starts from 20.6 kHz with a phase reversal occurring at 20.8 kHz. Prominent field dependent excitations are observed only in the frequency range of 0.1–0.5 THz, beyond which no significant response is observed (see inset). (b) Relative THz phase response plotted against the frequency of the applied external field. The relative phase response appears to exhibit a rippling surface-wave like behavior. The phase shift is apparent at 0.259 THz and increases gradually until 0.276 THz. A sudden phase reversal relative to that at 0.276 THz was observed at 0.285 THz along with an indication of maximum*

*response. For higher THz frequencies, the relative phase response flattens out and disappears completely beyond 0.302 THz. (c) Contour of the surface displacement profile measured with a vibrometer for excitation frequencies ranging between 19.2 kHz to 20.9 kHz. The surface displacement profile shows the maximum displacement of 108 pm at 20.7 kHz indicating it to be the regime of piezo-response along with rendering correspondence to (d), the THz phase response observed at 0.285 THz. Reproduced from reference 17 with permission from Springer Nature, Copyright 2016.*

When a device is operated at resonance condition, the materials possess two consecutive nodes: a resonant and anti-resonant node as shown in Figure 12. At resonant frequency ($f_r$), the system sees minimum value of impedance if the resistance caused by mechanical losses is not considered. Whereas at anti-resonant frequency ($f_a$), the impedance increases to maximum if resistance caused by the mechanical losses is ignored. However, when it is operated close to the resonance state, the phonon vibrations created are transformed into surface waves, which directly alter the material's refractive index. Because the incident THz waves encounter a different optical path, this will change the phase of reflected waves. The resultant phase shift can be controlled over the combined mechanism such as (i) change in path length due to the deformation in thickness of materials and thus index modulation, (ii) Polarization-mediated coupling occurs through a resonance-induced surface charge density gradient which effectively modulates the local field experienced by the THz waves.

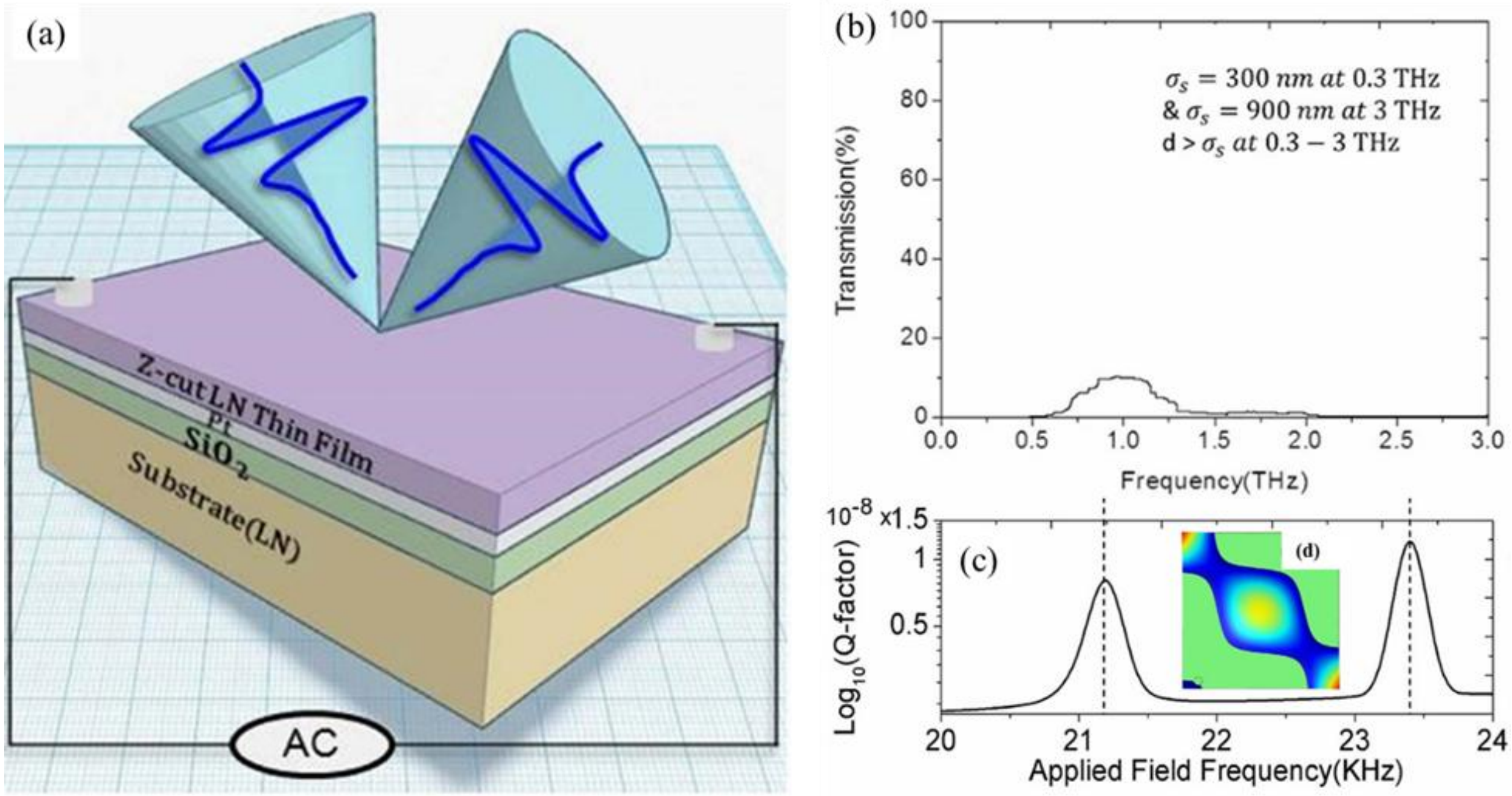

***Figure 11**. Experimental-setup, the crystallographic orientation and morphology of the multilayered single crystal LN thin film system employed for THz wave modulation. (a) Schematic of the multilayered thin film system comprising an ion-sliced single crystal z-cut LN (504 nm thick) deposited on platinized silica grown on an LN substrate (500 μ m thick). The top layer is excited by applying an external a.c. field via two silver electrodes deposited at opposite corners diagonally placed on the surface. (b) THz transmission spectra of*

*the platinized thin film system. As calculated from the skin depth $\sigma_s$ of platinum, the embedded 400 nm thick layer is opaque to THz waves resulting in a very low transmission amplitude therefore making the system suitable for reflection mode of operation where the loss of incident energy due to transmission is low. (c) quality factor plots. The insets are 3D representations of the surface displacements simulated at the two resonant conditions. (d) Simulated surface-contour displacement plot at 21.2 kHz. Reproduced from reference 17 with permission from Springer Nature, Copyright 2016.*

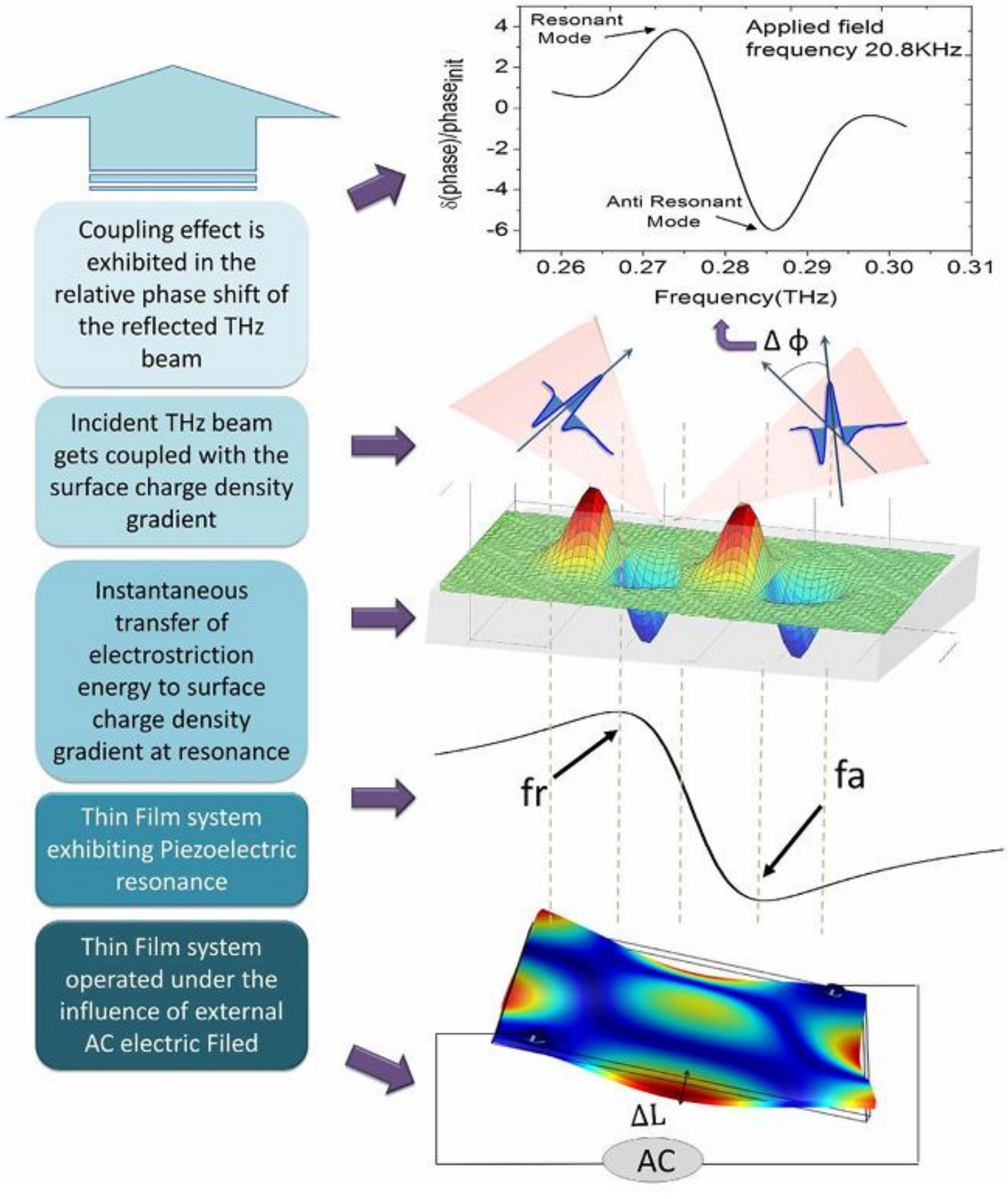

***Figure 12.*** *Resonance-defined electric field-controlled phase modulation of the incident THz beam. Schematic illustrations explain the interplay between electromagnetic (EM) and lattice-phonon vibrations at piezo-resonance in a ferroelectric material. When the thin film is operated at the resonance condition ($f_r$ and $f_a$ represents the resonant and the anti-resonant frequency nodes respectively) by applying a low frequency external electric field (with $\Delta L$ depicting the path-length difference, experienced by the reflected beam due to the applied electric field), the material density gradient gets instantaneously translated to the surface charge density gradient. The incident THz beam gets coupled with this surface charge density gradient to modulate its phase transfer function, which is prominently depicted in the relative phase change observed in the reflected beam. The resonance and anti-resonance nodes are distinctively captured in the relative phase shift $\Delta\phi$ when excited at the resonance condition i.e. at 20.8 kHz. Reproduced from reference 17 with permission from Springer Nature, Copyright 2016.*

THz modulation based on strain induced around the piezoelectric mode uses low frequency AC voltage (~ $10V_{pp}$). Consequently, the required driving electric field is relatively inexpensive compared to carrier-injection or high-speed electronic modulators that require GHz electronics or high bias voltages. The energy coupled via mechanical resonance allows strong optical path modulation with very small displacement, which helps in achieving low-power operation. Considering the device operation, it can be operated at 300 K without any cryogenic requirements while some electronics or quantum well based modulators require low temperatures.[17]

The strain driven approach enables efficient phase modulation of incident THz waves through tuning of refractive index, rather than relying on free-carrier dynamics or optical excitation. The key advantages are as follows. First, it enables low-power operation, as modulation can be achieved using relatively small AC voltages (on the order of ~10 Vpp at tens of kHz), significantly lower than the high bias fields or GHz-driving electronics. Second, the absence of complex architectures such as metamaterials or quantum-well heterostructures simplifies device fabrication and improves scalability. Third, operation at room temperature (~300 K) eliminates the need for cryogenic conditions often required in certain quantum or high-mobility electronic systems. Additionally, the resonance-enhanced electromechanical coupling allows strong optical path modulation with minimal mechanical displacement, improving electro-optic efficiency. This is particularly observed in organic nonlinear optical crystals, where low-frequency piezoelectric resonances (below 1 MHz) further reduce the required driving field. However, there are several limitations such as, the modulation speed is inherently constrained by the mechanical resonance frequency, which is typically in the kHz–MHz range and therefore slower than ultrafast optical or purely electronic THz modulators operating at GHz or higher frequencies. Furthermore, resonance-based operation may introduce bandwidth limitations and sensitivity to mechanical damping or environmental perturbations. Material stability, especially in organic systems, and long-term device reliability under cyclic strain also require careful consideration. Finally, while the approach reduces electrical complexity, precise control of mechanical resonance conditions is necessary to achieve consistent performance. Table 3. provides a comparative summary of major materials types used for THz modulation along with their corresponding modulation strategies, suitable application scenarios, and associated technical challenges.

***Table 3**: Comparative overview of various material systems, with corresponding modulation strategies and challenges.*

| **Material platform** | **Modulation strategy** | **Mechanism** | **Suitable applications** | **Main challenges** |
|---|---|---|---|---|
| Semiconductor materials (Si, GaAs, InSb) | Optical pumping / carrier modulation | Photoexcited free carriers modify conductivity and dielectric constant (Drude response) | High-speed THz communication, THz switches, spectroscopy | High optical pump power, carrier recombination losses, limited modulation depth |
| Metamaterials / metasurfaces | Resonance tuning / structural phase modulation | Artificial resonators manipulate THz phase, amplitude, and wavefront through LC resonance engineering | Beam steering, holography, adaptive optics, wavefront shaping | Narrow bandwidth, fabrication complexity, insertion loss |
| Graphene-based 2D materials | Electrical gating and optical excitation | Fermi-level tuning and ultrafast carrier dynamics alter THz conductivity | 6G communication, integrated THz photonics, ultrafast modulators | Limited interaction length, insertion loss, large-area fabrication difficulty |
| Transition metal dichalcogenides ($MoS_2$, $WSe_2$) | All-optical and strain-assisted modulation | Strong light–matter interaction and tunable bandgap modify THz transmission | Flexible THz devices, hybrid metasurfaces, tunable photonics | Interface defects, environmental instability, scalability |
| Phase-change materials ($VO_2$, GST) | Thermal/electrical phase-transition modulation | Insulator–metal or amorphous–crystalline transition changes dielectric properties | Reconfigurable THz optics, THz switches, adaptive metasurfaces | Thermal hysteresis, heat dissipation, slower switching |
| Liquid crystal modulators | Electric-field-driven birefringence modulation | Molecular reorientation changes refractive index and phase delay | THz phase shifters, beam steering, imaging systems | Slow response time, temperature sensitivity |
| Flexible substrate-based modulators | Mechanical deformation and optical modulation | Stretching/bending alters resonant geometry or dielectric response | Wearable sensors, flexible communication systems, biomedical devices | Mechanical fatigue, limited thermal stability |
| Organic nonlinear optical crystals (DAST, DSTMS, OH1, COANP) | Electro-optic and optical rectification modulation | Strong nonlinear susceptibility and electro-optic coefficients enable THz generation/modulation | Broadband THz generation, low-power phase modulation, spectroscopy | Moisture sensitivity, crystal fragility, thermal instability |
| Strain-driven piezoelectric crystals ($LiNbO_3$, MBA-NP, semi-organic crystals) | Piezoelectric resonance-based strain modulation | Converse piezoelectric effect induces strain and refractive-index modulation at resonance | Low-power THz phase modulators, sensing, adaptive THz optics | Limited modulation speed, resonance-dependent bandwidth, mechanical damping |
| Piezoelectric MEMS metasurfaces | Electromechanical reconfiguration | MEMS actuation changes resonator geometry and phase response | Programmable metasurfaces, beam steering, reconfigurable THz systems | Pull-in instability, mechanical fatigue, slower response |

## 7. Applications of Tunable THz Phase Modulators

### 7.1. Terahertz Wireless Communication Systems

Tunable phase modulation in the THz frequency range has emerged as a key enabling technology for next-generation wireless communication systems. The THz band (0.1–10 THz) offers extremely large spectral bandwidth compared with microwave and millimeter-wave frequencies, making it attractive for ultra-high-speed data transmission. In many communication schemes, information is encoded in the phase of the carrier wave through techniques such as phase shift keying (PSK) and quadrature phase shift keying (QPSK). Tunable phase modulators allow dynamic control of the phase of THz radiation, enabling efficient signal encoding and adaptive communication protocols. Active phase manipulation is particularly important for the development of future 6G wireless networks, high-capacity indoor communication links, and chip-to-chip communication systems. Recent studies have shown that tunable THz photonic components based on metamaterials and phase-transition materials can provide broadband modulation capabilities required for high-speed communication devices.[84]

### 7.2. Terahertz Imaging and Security Screening

Another significant application of tunable THz phase modulation lies in THz imaging systems. THz radiation is non-ionizing and can penetrate many non-conductive materials such as plastics, fabrics, ceramics, and paper. These properties make THz waves highly suitable for security screening, biomedical diagnostics, and industrial inspection. Phase-sensitive imaging techniques use variations in the phase of THz waves to extract information about internal structures and material boundaries. Tunable phase modulators improve image quality by enabling phase contrast imaging and depth-resolved imaging. This capability allows detection of concealed objects, structural defects, or hidden features in layered materials. Therefore, increasing the spatial resolution and sensitivity of sophisticated THz imaging systems used in security detection and non-destructive testing requires the capacity to precisely regulate the phase of THz radiation. [85].

### 7.3. Terahertz Spectroscopy and Material Characterization

Tunable THz phase modulation plays a crucial role in THz spectroscopy, especially in THz time-domain spectroscopy (THz-TDS). In THz-TDS systems, both the amplitude and phase of the transmitted or reflected THz pulse are measured to determine material parameters such as refractive index, dielectric constant, and absorption coefficient. Phase-sensitive measurements are particularly important because they provide accurate information about the propagation of THz waves through materials. By controlling the phase of the THz signal, tunable phase modulators enable enhanced spectral resolution and improved signal-to-noise ratio in spectroscopic measurements. This technique has widely been used to study molecular vibrations, phonon resonances, and carrier dynamics in semiconductors and complex materials. Consequently, THz spectroscopy has found applications in chemical sensing, pharmaceutical analysis, explosive detection, and characterization of advanced electronic and photonic materials [86].

### 7.4. Wavefront Manipulation and Metasurface Devices

In recent years, metasurfaces have emerged as an effective platform for controlling the phase of THz waves. These engineered surfaces consist of subwavelength structures that can impose spatially varying phase shifts on incident electromagnetic waves. By dynamically tuning the phase response of each unit cell, tunable metasurfaces can manipulate the propagation direction, focusing behavior, and polarization state of THz radiation. Such devices enable functionalities including beam steering, metalenses, and vortex beam generation. Incorporating phase-change materials or graphene into metasurfaces allows active control of THz wavefronts, enabling reconfigurable optical devices. This technology is expected to play an important role in future THz photonic systems, including adaptive imaging systems and compact optical components [87].

### 7.5. Terahertz Holography

Holography is another field where tunable THz phase modulation has significant impact. In holographic systems, the reconstruction of three-dimensional images relies on precise control of the phase distribution of electromagnetic waves. By modulating the phase of THz waves across a spatial light modulator or metasurface, it is possible to generate complex holographic patterns and reconstruct high-resolution images. Tunable phase modulators enable real-time reconfiguration of holographic patterns, making them suitable for dynamic imaging systems and information processing. The capability of local phase control of electromagnetic waves forms the foundation for holographic imaging and advanced wavefront engineering technologies in the THz regime [23].

### 7.6. Non-Destructive Testing and Industrial Inspection

THz radiation has become an important tool for non-destructive testing (NDT) because it can probe the internal structure of materials without causing damage. Phase-sensitive THz measurements provide additional information about optical path differences inside multilayer materials. Tunable phase modulators enhance the detection of structural defects such as voids, cracks, and delamination in composite materials. These techniques are particularly useful in industries such as aerospace, semiconductor manufacturing, and pharmaceutical quality control. In layered structures, phase information allows accurate measurement of layer thickness and material interfaces, which is essential for evaluating the structural integrity of advanced materials and electronic components [85].

### 7.7. Terahertz Radar and Sensing Technologies

The extremely short wavelength of THz radiation enables very high spatial resolution in radar and sensing applications. Tunable phase modulation plays an essential role in beamforming and phased-array radar systems operating in the THz frequency range. By controlling the phase of multiple antenna elements, the direction of the radiated beam can be dynamically adjusted without mechanical movement. This capability is essential for applications such as autonomous vehicle navigation, drone guidance, and high-resolution environmental monitoring. Additionally, THz radar systems can be used for remote sensing and detection of small objects due to their high sensitivity to surface features and material properties [42].

## 8. Future Outlook

The rapid development of THz phase modulators has demonstrated the enormous potential of externally stimulated materials for dynamic manipulation of THz wave amplitude and phase. Electrical, optical, thermal, and mechanical approaches for modulation have shown distinct advantages; however, significant challenges remain in achieving simultaneously large phase modulation depth, ultrafast response, low insertion loss, low power consumption, and long-term operational stability. Although liquid-crystal-based devices can provide nearly 360° phase tuning, their relatively slow response limits high-speed applications. In contrast, optically pumped THz modulators offer ultrafast operation but often require complicated optical excitation systems and typically exhibit limited phase tuning ranges. Consequently, most currently reported THz phase modulators remain at the initial stage because of the complexity in fabrication, environmental instability, and material degradation issues.

Among the emerging modulation mechanisms, strain-engineered piezoelectric THz modulators offer a particularly promising pathway for next-generation THz technologies. In piezoelectric systems, externally applied electric fields generate controllable mechanical deformation through the converse piezoelectric effect, thereby modifying the optical path length, dielectric permittivity, and resonance characteristics of the active medium. These strain-induced variations in polarization and surface charge density gradients strongly influence the phase transfer function of THz waves, especially in resonant metamaterial architectures. Compared with purely electrical or optical modulation schemes, piezoelectric strain engineering provides several important advantages, including low driving power, precise electrical tunability, room-temperature operation, compact integration capability, and compatibility with reconfigurable metasurface platforms.

An integration of piezoelectric thin films with two-dimensional (2D) materials such as $MoS_2$ and h-BN with heterogeneous stacking remains a future direction in the field of THz modulation. Owing to their atomic-scale thickness, excellent carrier mobility, strong strain sensitivity, 2D materials can significantly enhance electromechanical coupling at the nanoscale. Recent studies have shown that strain-induced bandgap tuning and piezo-phototronic effects in $MoS_2$ and related transition metal dichalcogenides can effectively modulate electronic and optical responses under external mechanical deformation. Integrating these materials with piezoelectric substrates or metamaterial resonators could enable ultrathin THz modulators with enhanced phase tunability, reduced insertion loss, and improved modulation speed. Furthermore, van der Waals heterostructures combining piezoelectric films with graphene, $MoS_2$, and h-BN may allow dynamic control of carrier concentration, plasmonic resonance, and local electromagnetic confinement in THz metasurfaces.

Another major gap in current THz phase modulation research is the absence of a comprehensive theoretical framework describing phase-matching conditions between piezoelectric resonance frequencies and THz spectral responses. Future research should focus on developing multiphysics theoretical models that couple electromechanical dynamics, phonon interactions, dielectric dispersion, and THz electromagnetic propagation. Such models should account for strain-induced anisotropic dielectric modulation, nonlinear electromechanical coupling, and resonance synchronization between piezoelectric oscillations and THz metamaterial modes.

Organic and semi-organic nonlinear optical (NLO) crystals also represent an attractive yet relatively underexplored route for low-power room-temperature THz modulation. Compared with conventional inorganic piezoelectric materials, organic NLO crystals often exhibit exceptionally large electro-optic coefficients, lower dielectric constants, lightweight structures, and mechanical flexibility. Materials such as DAST (4-N,N-dimethylamino-4′-N′-methylstilbazolium tosylate), OH1, BNA, and related semi-organic crystals have demonstrated strong THz generation and electro-optic responses. However, their practical implementation in THz modulators remains limited by crystal fragility, moisture sensitivity, thermal instability, and fabrication difficulties for large-area devices. Future work should focus on improving crystal growth techniques, enhancing environmental stability through encapsulation strategies, and developing hybrid organic–inorganic composite architectures that combine strong nonlinear responses with improved mechanical stability. Integration of organic NLO crystals with strain-tunable piezoelectric platforms may further enable low-voltage THz phase modulation with reduced power consumption and enhanced operational bandwidth. An accelerating development in this area and establishing advanced characterization techniques to achieve a clear understanding of the structural influences on modulation behavior are immediate demands in this direction. In addition, challenges such as mechanical stability and limited operating frequency must be addressed to enable practical THz modulation applications. Furthermore, research should be directed on numerous existing organic and inorganic nonlinear optical materials with high piezoelectric coefficients, including identified materials this this review, as this offers significant potential for efficient phase modulation.

Flexible and wearable THz modulators are expected to become increasingly important for next-generation sensing, biomedical diagnostics, soft robotics, and human–machine interface technologies. The combination of piezoelectric thin films with flexible polymer substrates offers exciting opportunities for mechanically deformable THz devices. Flexible THz modulators could enable real-time physiological monitoring, wearable spectroscopy systems, and adaptive wireless communication platforms. Materials such as PVDF, P(VDF-TrFE), and ultrathin AlN or ScAlN piezoelectric films are particularly attractive due to their mechanical flexibility and compatibility with low-temperature fabrication processes. Eventhough, maintaining stable THz performance under repeated bending, stretching, and cyclic mechanical loading remains a major challenge.

Artificial intelligence and machine learning (ML) are also expected to transform the design and optimization of THz piezoelectric materials and devices. Conventional trial-and-error material development approaches are often time-consuming because piezoelectric performance depends on highly coupled parameters such as crystal orientation, defect density, alloy composition, domain structure, and electromechanical coupling coefficients. ML-assisted inverse design frameworks can rapidly explore multidimensional parameter spaces and predict optimal material compositions for enhanced THz modulation performance. Specifically, data-driven methods in conjunction with finite-element analysis, density functional theory (DFT), and high-throughput materials screening may accelerate the discovery of novel piezoelectric alloys and hybrid heterostructures with superior electromechanical coupling.

Scalability and complementary metal-oxide-semiconductor (CMOS) compatibility remain critical barriers preventing the commercialization of piezoelectric THz modulators. Most reported devices rely on complex fabrication methods, non-standard substrates, or fragile laboratory-scale assembly processes that are difficult to integrate into existing semiconductor manufacturing lines. Future efforts should therefore prioritize CMOS-compatible piezoelectric materials such as AlN, ScAlN, $HfO_2$-based ferroelectrics, and low-temperature deposited thin films that can be directly integrated with silicon photonic and electronic platforms. Advances in wafer-scale nanofabrication, heterogeneous integration, and low-loss packaging technologies will be essential for realizing compact, reliable, and mass-producible THz systems.

Overall, future THz phase modulators are expected to evolve toward multifunctional, intelligent, low-power, and highly integrated platforms that combine piezoelectric strain engineering, metamaterials, 2D materials, organic nonlinear crystals, and AI-assisted optimization. Continued progress in material science, theoretical modeling, scalable fabrication, and system-level integration will be essential for translating laboratory-scale demonstrations into commercially viable THz technologies for adaptive photonic systems, imaging, sensing, and next-generation wireless communication.

## AUTHOR INFORMATION

### Corresponding Author

**Rajaboopathi Mani** – Key Laboratory of Emergent Materials (KLEM), Department of Physics and Nanotechnology, SRM Institute of Science and Technology, Kattankulathur - 603203 Chengalpattu, Tamil Nadu, India. orcid.org/0000-0002-2535-8987; Email: rajaboom@srmist.edu.in

### Author

**Aarifa Muhammed Ashraf** – Key Laboratory of Emergent Materials (KLEM), Department of Physics and Nanotechnology, SRM Institute of Science and Technology, Kattankulathur - 603203, Chengalpattu, Tamil Nadu, India. orcid.org/0009-0003-5987-7579

### Author Contributions

Aarifa Muhammed Ashraf: Writing – original draft preparation, Visualization, Rajaboopathi Mani: Writing – review & editing, Supervision, Project administration, Conceptualization.

### Funding

This research was supported by the Selective Excellence Research Initiative, SRMIST for their funding support (SRMIST/R/AR(A)/SERI2024/ 174/48).

### Notes

The authors declare no competing financial interest.

### Conflict of Interest

The corresponding author has previously published research articles related to the topics discussed in this review. These publications have been appropriately cited where relevant. The authors declare that every effort was made to ensure an objective, balanced, and comprehensive evaluation of the literature presented in this manuscript. Therefore, editors and readers can assess potential bias accordingly.

## ACKNOWLEDGEMENT

The author AMA express her sincerest gratitude to SRM Institute of Science and Technology, Tamil Nadu, India for granting institutional fellowship.

## References

1 H. Barhum, C. McDonnell, T. Alon, R. Hammad, M. Attrash, T. Ellenbogen and P. Ginzburg, *ACS Appl. Mater. Interfaces*, 2023, **15**, 8590–8600.

2 L. Mutter, F. D. Brunner, Z. Yang, M. Jazbinšek and P. Günter, *J. Opt. Soc. Am. B*, 2007, **24**, 2556.

3 Y. Lu, Q. Zhang, Q. Wu, Z. Chen, X. Liu and J. Xu, *Nat. Commun,* 2021, **12**, 3183.

4 Y. Lu, Y. Huang, J. Cheng, R. Ma, X. Xu, Y. Zang, Q. Wu and J. Xu, *Nanophotonics*, 2024, **13**, 3279–3298.

5 Z. Ren, J. Xu, J. Liu, B. Li, C. Zhou and Z. Sheng, *ACS Appl. Mater. Interfaces*, 2022, **14**, 26923-26930.

6 K. Kawase, Y. Ogawa, Y. Watanabe and H. Inoue, *Opt. Express*, 2003, **11**, 2549.

7 S. Koenig, D. Lopez-Diaz, J. Antes, F. Boes, R. Henneberger, A. Leuther, A. Tessmann, R. Schmogrow, D. Hillerkuss, R. Palmer, T. Zwick, C. Koos, W. Freude, O. Ambacher, J. Leuthold and I. Kallfass, *Nat Photonics,* 2013, **7**, 977–981.

8 W. J. Choi, M. R. Armstrong, J. H. Yoo and T. Lee, *J. Mater. Chem. C Mater.*, 2024, **12**, 9002–9011.

9 M. Welsch, A. Singh, S. Winnerl, A. Pashkin, M. Xu, M. Li, M. Helm and H. Schneider, *J. Infrared, Millimeter, Terahertz Waves,* 2021, **42**, 537–546.

10 X. L. Wu, S. J. Xiong, Z. Liu, J. Chen, J. C. Shen, T. H. Li, P. H. Wu and P. K. Chu, *Nat. Nanotechnol,* 2011, **6**, 103–106.

11 Z. T. Ma, Z. X. Geng, Z. Y. Fan, J. Liu and H. D. Chen, *Research*, 2019, **2019**, 6482975.

12 M. Rahm, J. S. Li and W. J. Padilla, *J. Infrared, Millimeter, Terahertz Waves,* 2012, **34**, 1–27.

13 J. Zhou, D. R. Chowdhury, R. Zhao, A. K. Azad, H. T. Chen, C. M. Soukoulis, A. J. Taylor and J. F. O'Hara, *Phys. Rev. B*, 2012, **86**, 035448.

14 J. Li, *Opt. Commun.*, 2007, **269**, 98–101.

15 B. Sun, X. He, J. Jiang, Y. Yao and G. Lu, *Opt. Laser Technol.*, 2024, **171**, 110342.

16 H. Zeng, S. Gong, L. Wang, T. Zhou, Y. Zhang, F. Lan, X. Cong, L. Wang, T. Song, Y. C. Zhao, Z. Yang and D. M. Mittleman, *Nanophotonics*, 2022, **11**, 415–437.

17 M. Dutta, S. Betal, X. G. Peralta, A. S. Bhalla and R. Guo, *Sci. Rep.*, 2016, **6**, 38041.

18 J. Xu, Z. Ren, F. Su, J. Liu, B. Li and Z. Sheng, *ACS Appl. Electron. Mater.*, 2022, **4**, 6269–6274.

19 R. Kersting, G. Strasser and K. Unterrainer, *Electron. Lett.*, 2000, **36**, 1156–1158.

20 H. T. Chen, W. J. Padilla, M. J. Cich, A. K. Azad, R. D. Averitt and A. J. Taylor, *Nat. Photonics*, 2009, **3**, 148–151.

21 R. Degl'innocenti, H. Lin and M. Navarro-Cía, *Nanophotonics*, 2022, **11**, 1485–1514.

22 S. H. Lee, M. Choi, T. T. Kim, S. Lee, M. Liu, X. Yin, H. K. Choi, S. S. Lee, C. G. Choi, S. Y. Choi, X. Zhang and B. Min, *Nat. Mater.,* 2012, **11**, 936–941.

23 Z. Miao, Q. Wu, X. Li, Q. He, K. Ding, Z. An, Y. Zhang and L. Zhou, *Phys. Rev. X*, 2015, **5**, 041027.

24 R. Yan, S. Arezoomandan, B. Sensale-Rodriguez and H. G. Xing, *ACS Photonics*, 2016, **3**, 315–323.

25 Z. Wang, J. Qiao, S. Zhao, S. Wang, C. He, X. Tao and S. Wang, *InfoMat*, 2021, **3**, 1110–1133.

26 H. Zeng, H. Liang, Y. Zhang, L. Wang, S. Liang, S. Gong, Z. Li, Z. Yang, X. Zhang, F. Lan, Z. Feng, Y. Gong, Z. Yang and D. M. Mittleman, *Nat. Photonics*, 2021, **15**, 751–757.

27 J. Zheng, X. Zhang, L. Liu, Q. Li, L. Singh, J. Han, F. Yan and W. Zhang, *J. Infrared, Millimeter, Terahertz Waves*, 2017, **38**, 1034–1046.

28 Y. Ji, X. Jiang, F. Fan, H. Zhao, J. Cheng, X. Wang, S. Chang, Y. Ji, X. Jiang, F. Fan, H. Zhao, J. Cheng, X. Wang, S. Chang, *OExpr*, 2023, **31**, 1269.

29 C. Xu, Z. Ren, J. Wei and C. Lee, *iScience*, 2022, **25**, 103799.

30 J. Tebart, J. Dittmer, T. Haddad, P. Matalla, P. Lu, S. Randel and A. Stöhr, *Int. J. Microw. Wirel. Technol.*, 2025, **17**, 276–284.

31 J. H. Chen, B. C. Zheng, G. H. Shao, S. J. Ge, F. Xu and Y. Q. Lu, *Light: Sci. Appl.,* 2015, **4**, e360–e360.

32 J. Qiao, S. Wang, Z. Wang, C. He, S. Zhao, X. Xiong, S. Wang, X. Zhang and X. Tao, *Adv. Opt. Mater.*, 2020, **8**, 2000160.

33 W. Li, B. Chen, C. Meng, W. Fang, Y. Xiao, X. Li, Z. Hu, Y. Xu, L. Tong, H. Wang, W. Liu, J. Bao and Y. R. Shen, Nano Lett., 2014, **50**, 1925–1928.

34 J.-S. Li and M.-S. Hu, *Sci. Rep*.,2020, **10**, 6605.

35 G. Liang, X. Hu, X. Yu, Y. Shen, L. H. Li, A. G. Davies, E. H. Linfield, H. K. Liang, Y. Zhang, S. F. Yu and Q. J. Wang, *ACS Photonics*, 2015, **2**, 1559–1566.

36 B. Sensale-Rodriguez, R. Yan, M. M. Kelly, T. Fang, K. Tahy, W. S. Hwang, D. Jena, L. Liu and H. G. Xing, *Nat. Commun.,* 2012, **3**, 780.

37 M. Alaloul and M. Rasras, *ACS Omega*, 2021, **6**, 7576–7584.

38 P. Mishra, V. Srivastava and Sunny, *2023 IEEE Workshop on Recent Advances in Photonics, WRAP 2023, 2023,1–3.*

39 H. Yu, B. Zhu, X. Qi and L. Guo, *Semiconductors*, 2024, **58**, 558–564.

40 G. Wang, M. Chu, J. Jiang, Y. Jia, J. Liu, H. Yu, H. Meng and W. Shi, *Phys. Chem. Chem. Phys*, 2026, **28**, 3423–3433.

41 A. K. Sahoo, C. S. Yang, C. L. Yen, H. C. Lin, Y. J. Wang, Y. H. Lin, O. Wada and C. L. Pan, *Appl. Sci. 2019,* **9**, 761.

42 X. Liu, D. Yu, C. Sun, Z. Mei, H. Chen, K. Zheng, P. Yan, F. Rao, J. Xu, J. Qu and Y. Sun, *ACS Photonics*, 2024, **11**, 2595–2603.

43 Y. Dong, D. Yu, G. Li, M. Lin and L. A. Bian, *Symmetry,* 2021, **13**, 2230.

44 S. Liu, J. Feng, Y. Tian, H. Zhao, L. Jin, B. Ouyang, J. Zhu and J. Guo, *Front. Optoelectron.,* 2022, **15**, 9.

45 H. Tao, A. C. Strikwerda, K. Fan, W. J. Padilla, X. Zhang and R. D. Averitt, *Phys. Rev. Lett.*, 2009, **103**, 147401.

46 M. Manjappa, P. Pitchappa, N. Singh, N. Wang, N. I. Zheludev, C. Lee and R. Singh, *Nat. Commun.*, 2018, **9**, 4056.

47 H. Ma, X. Xiao, Y. Wang, Y. Sun, B. Wang, X. Gao, E. Wang, K. Jiang, K. Liu and X. Zhang, *Sci. Adv.*, 2021, **7**, 3438.

48 H. Ma, Y. Wang, R. Lu, F. Tan, Y. Fu, G. Wang, D. Wang, K. Liu, S. Fan, K. Jiang and X. Zhang, *J. Mater. Chem. C Mater.*, 2020, **8**, 10213–10220.

49 H. F. Zhu, L. H. Du, J. Li, Q. W. Shi, B. Peng, Z. R. Li, W. X. Huang and L. G. Zhu, *Appl. Phys. Lett.*, 2018, **112**.

50 Z. Ren, L. Cheng, L. Hu, C. Liu, C. Jiang, S. Yang, Z. Ma, C. Zhou, H. Wang, X. Zhu, Y. Sun and Z. Sheng, *ACS Appl. Mater. Interfaces*, 2020, **12**, 48811–48819.

51 H. Cai, S. Chen, C. Zou, Q. Huang, Y. Liu, X. Hu, Z. Fu, Y. Zhao, H. He and Y. Lu, *Adv. Opt. Mater.*, 2018, **6**, 1800257.

52 W. Lai, R. Shi, H. Yuan, G. Liu, A. Amini, C. Cheng and W. Lai, *ACS Appl. Electron. Mater.*, 2021, **3**, 3044–3051.

53 W. Liang, Y. Jiang, J. Guo, N. Li, W. Qiu, H. Yang, Y. Ji and S. N. Luo, *Adv. Opt. Mater.*, 2019, **7**, 1900647.

54 H. Zhu, J. Li, L. Du, W. Huang, J. Liu, J. Zhou, Y. Chen, S. Das, Q. Shi, L. Zhu and C. Liu, *Adv. Mater. Interfaces*, 2020, **7**, 2001297.

55 A. V. Ivanov, A. Y. Tatarenko, A. A. Gorodetsky, O. N. Makarevich, M. 5-Cía, A. M. Makarevich, A. R. Kaul, A. A. Eliseev and O. V. Boytsova, *ACS Appl. Nano Mater.*, 2021, **4**, 10592–10600.

56 M. Tanaka, F. Miyamaru, M. Hangyo, T. Tanaka, M. Akazawa and E. Sano, *Opt. Lett.,* 2005, **30**, 1210–1212.

57 J. sheng Li and J. quan Yao, *Opt. Commun.*, 2008, **281**, 5697–5700.

58 L. Fekete, F. Kadlec, P. Kužel and H. Němec, *Opt. Lett.*, 2007, **32**, 680–682.

59 L. Deng, J. Teng, H. Liu, Q. Y. Wu, J. Tang, X. Zhang, S. A. Maier, K. P. Lim, C. Y. Ngo, S. F. Yoon and S. J. Chua, *Adv. Opt. Mater.*, 2013, **1**, 128–132.

60 K. S. Novoselov, A. K. Geim, S. V. Morozov, D. Jiang, Y. Zhang, S. V. Dubonos, I. V. Grigorieva and A. A. Firsov, *Science*, 2004, **306**, 666–669.

61 P. Y. Chen and A. Alu, *IEEE Trans. Terahertz Sci. Technol.*, 2013, **3**, 748–756.

62 M. Liu, X. Yin, E. Ulin-Avila, B. Geng, T. Zentgraf, L. Ju, F. Wang and X. Zhang, *Nature, 2011*, **474**, 64–67.

63 P. Weis, J. L. Garcia-Pomar, M. Höh, B. Reinhard, A. Brodyanski and M. Rahm, *ACS Nano*, 2012, **6**, 9118–9124.

64 X. Zou, J. Shang, J. Leaw, Z. Luo, L. Luo, C. La-O-Vorakiat, L. Cheng, S. A. Cheong, H. Su, *Phys. Rev. Lett.*, 2013, **110**, 067401.

65 W. Zheng, F. Fan, M. Chen, S. Chen and S. J. Chang, *AIP Adv.*,2016, **6**, 075105.

66 Y. Yoon, Z. Lu, C. Uzundal, R. Qi, W. Zhao, S. Chen, Q. Feng, W. Kim, M. H. Naik, K. Watanabe, T. Taniguchi, S. G. Louie, M. F. Crommie and F. Wang, *Nature*, 2024, **631**, 771–776.

67 Y. Cheng, X. Lin, C. Wang, Z. Zhang, C. Wang, L. Geng, P. Suo, J. Du and G. Ma, *J. Phys. Chem. Lett.*, 2025, **16**, 2654–2660.

68 T. Kleine-Ostmann and T. Nagatsuma, *J. Infrared, Millimeter, Terahertz Waves*, 2011, **32**, 143–171.

69 D. Y. Khang, H. Jiang, Y. Huang and J. A. Rogers, *Science*, 2006, **311**, 208–212.

70 K. Fan, X. Zhao, J. Zhang, K. Geng, G. R. Keiser, H. R. Seren, G. D. Metcalfe, M. Wraback, X. Zhang and R. D. Averitt, *IEEE Trans. Terahertz Sci. Technol.*, 2013, **3**, 702–708.

71 F. Pan, G. Knöpfle, C. Bosshard, S. Follonier, R. Spreiter, M. S. Wong and P. Günter, *Appl. Phys. Lett.*, 1996, **69**, 13–15.

72 C. Hunziker, S.-J. Kwon, H. Figi, F. Juvalta, O.-P. Kwon, M. Jazbinsek and P. Günter, *J. Opt. Soc. Am. B*, 2008, **25**, 1678.

73 Ch. Bosshard, G. Chapuis, K. Sutter and P. Günter, *J. Opt. Soc. Am. B*, 1989, **6**, 721.

74 M. V. Shankar, K. B. R. Varma, R. T. Bailey, F. R. Cruickshank, D. Pugh and J. N. Sherwood, *J. Appl. Phys.*, 1997, **81**, 2370–2374.

75 V. Krishnakumar, M. Rajaboopathi and R. Nagalakshmi, *Physica B Condens. Matter*, 2012, **407**, 1119–1123.

76 C.-M. Tu, L.-H. Chou, Y.-C. Chen, P. Huang, M. Rajaboopathi, C.-W. Luo, K.-H. Wu, V. Krishnakumar, T. Kobayashi, *Opt Express*, 2016, **24**, 5039–5044.

77 R. K. Saripalli, T. Chakraborty, H. L. Bhat and S. Elizabeth, *Appl. Phys. A*, 2016, **122**, 438.

78 S. Brahadeeswaran, H. L. Bhat, N. S. Kini, A. M. Umarji, P. Balaya and P. S. Goyal, *J. Appl. Phys.*, 2000, **88**, 5935–5940.

79 P. Kerkoc, S. Maruo, S. Horinouchi and K. Sasaki, *Appl. Phys. Lett.*, 1999, **74**, 3105–3106.

80 K. Raghavendra Rao, H. L. Bhat and S. Elizabeth, *Mater. Chem. Phys.*, 2013, **137**, 756–763.

81 C. Ranjith Dev Inbaseelan, P. Sakthivel and S. Saravanan, *Optik*, 2016, **127**, 2720–2724.

82 P. V. Lenzo, E. G. Spencer and K. Nassau, *JOSA,* 1966, **56**, 633–635.

83 F. B. Ahr, H. Deutsches Elektronen-Synchrotron and I. und N. Hamburg Univ. Fakultaet fuer Mathematik, 2017.

84 T. Nagatsuma, G. Ducournau and C. C. Renaud, *Nat. Photonics*, 2016, **10**, 371–379.

85 J. Wu, R. Liu, K. Cheng, Z. Zhang, D. Wang, X. Wang, Y. Liu, Y. Han and Y. Li, *Opt. Commun.*, 2025, **578**, 131481.

86 P. U. Jepsen, D. G. Cooke and M. Koch, *Laser Photon. Rev.*, 2011, **5**, 124–166.

87 M. Zhang, P. Dong, Y. Wang, B. Wang, L. Yang, R. Wu, W. Hou and J. Zhang, *Nanomaterials 2022*, 2022, **12**, 3592.